\input amstex
\documentstyle{amsppt}




\loadbold
\baselineskip=15 pt
\TagsOnRight

\define\KK{\Cal K}
\define\HH{\Cal H}
\define\BB{\Cal B}
\define\II{\Cal I}
\define\DD{\Cal D}
\define\TT{\Cal T}
\redefine\SS{\Cal S}
\define\UU{\Cal U}
\define\XX{\Cal X}

\define\EE{\Cal E}
\define\MM{\Cal M} 
\redefine\AA{\Cal A} 
\define\NN{\Cal N}
\define\BC{\Bbb C}
\define\BN{\Bbb N}
\define\BZ{\Bbb Z}
\define\BR{\Bbb R}
\define\BE{\Bbb E}
\define\BP{\Bbb P}
\define\half{{1\over2}}
\define\const{\text{const}}

\define\pr{\partial}
\define\e{\text{e}}
\define\Spec{\operatorname{Spec}}
\define\Dom{\operatorname{Dom}}
\define\Ran{\operatorname{Ran}}
\define\Lip{\operatorname{Lip}}
\define\ad{\operatorname{ad}}
\define\diam{\operatorname{diam}}
\define\w{\omega}
\define\bt{\beta}
\define\la{\lambda}
\define\al{\alpha}

\define\veps{\varepsilon}
\define\Ga{\Gamma}
\define\ga{\gamma}
\define\sg{\sigma}
\define\dl{\delta}
\define\vp{\varphi}

\define\gG{\goth{G}}
\define\gL{\goth{L}}

\define\hK{\hat K}
\define\hV{\hat V}
\define\tpsi{\tilde\psi}
\define\tG{\tilde G}
\define\tla{\tilde\lambda}
\define\tD{\tilde{\Cal D}}
\redefine\and{\text{and}}
\define\for{\text{for}}
\define\fora{\text{for all}}

\define\Vt{V(t)}
\define\BZN{\BZ\times\BN}
\define\fb{f(\beta)}
\define\Fb{F(\beta)}
\define\GO{\Gamma_0}
\define\Gl{\Gamma_\lambda}
\define\bl{(\beta,\lambda)}
\define\Gbl{\Gamma(\beta,\lambda)}
\define\of{^{\text{off}}}
\define\di{^{\text{diag}}}
\define\en{^{(\nu)} }
\define\em{^{(\mu)} }
\define\Lp{Lipschitz }
\define\tggap{2.1}
\define\tgV{2.2}
\define\tghat{2.3}
\define\tgnorm{2.4}
\define\tglam{2.5}
\define\tgel{2.6}
\define\tgsetIa{3.2}
\define\tgsetIb{3.3}
\define\tgGO{3.5}
\define\tgGla{3.6}
\define\tgexpo{3.9}
\define\tgOmega{3.10}
\define\tgdiow{3.11}
\define\tgpsi{3.12}
\define\tgeqI{4.5}
\define\tgeqII{4.6} 
\define\tgeqIII{4.7}
\define\tgvectors{4.8}
\define\tggMa{4.9a}
\define\tggMb{4.9b}
\define\tggMc{4.9c}
\define\tggM{4.9}
\define\tglaMa{4.10a}
\define\tglaMb{4.10b}
\define\tglaMc{4.10c}
\define\tglaM{4.10}
\define\tgS{5.1}
\define\tgineq{5.2}
\define\tgdistI{5.4} 
\define\tgdistII{5.5}
\define\tgbound{5.6}
\define\tgadbound{5.8}
\define\tgD{5.11}
\define\tgdecay{5.12}
\define\tgsimple{6.1}
\define\tgsetIIa{6.2}
\define\tgsetIIb{6.3} 
\define\tgW{6.4} 
\define\tgdomI{6.5} 
\define\tgWV{6.6}
\define\tgred{6.7}
\define\tgdiobl{6.9}
\define\tgL{6.10}
\define\tgdioop{6.11}
\define\tgadW{6.12}
\define\tgdomr{6.15}
\define\tgsols{6.16} 
\define\tgsol{6.17} 
\define\tgsetIIbb{6.18}
\define\tgvn{7.1}
\define\tgbasic{7.2}
\define\tgtwn{7.4}
\define\tgwn{7.5}
\define\tgdiof{7.7}
\define\tgdomII{7.8}
\define\tgexpand{7.9} 
\define\tgDD{7.10}
\define\tgIphi{7.11}
\define\tgalmost{8.1}
\redefine\tgG{8.2} 
\define\tgimpl{8.3} 
\define\tgderG{8.4}
\define\tgimplex{8.5} 
\define\tgsolvex{8.6}
\define\tgderla{8.7} 
\define\tgderlaII{8.8}  

\define\Arnold{1}
\define\Bellissard{2}
\define\BleherJauslinLebowitz{3}
\define\Chung{4}
\define\Combescure{5}
\define\DS1{6}
\define\DSV1{7}
\define\Eliasson{8}
\define\EnssVeselic{9}
\define\HowlandA{10}
\define\HowlandB{11}
\define\Joye{12}
\define\Kato{13}
\define\Kolmogorov{14}
\define\Messiah{15}
\define\Moser{16}
\define\Nenciu{17}
\define\OGC{18}
\define\ReedSimon{19}
\define\Rellich{20}
\define\Sambe{21}
\define\Siegel{22}
\define\Stein{23}
\define\Yajima{24}

\topmatter
\title
Perturbation of an eigen-value from a dense point
spectrum: a general Floquet Hamiltonian
\endtitle
\author P. Duclos$^{1,2}$, P. \v{S}\v{t}ov\'{\i}\v{c}ek$^3$
and M. Vittot$^1$
\endauthor
\leftheadtext{P. Duclos, P. \v{S}\v{t}ov\'{\i}\v{c}ek and M. Vittot}
\rightheadtext{Perturbation of an eigen-value for Floquet Hamiltonians}
\affil
$^1$ Centre de Physique Th\'eorique, CNRS-Luminy, Case 907\\
F-13288 Marseille cedex 9, France\\
(Unit\'e Propre de Recherche 7061)\\
$ $\\
$^2$ PhyMaT, Universit\' e de Toulon et du Var, BP 132\\
F-83957 La Garde cedex, France\\
$ $\\
$^3$ Department of Mathematics\\
Faculty of Nuclear Science, CTU\\
Trojanova 13, 120 00 Prague, Czech Republic\\
$ $\\
\endaffil

\thanks duclos\@naxos.unice.fr \endthanks
\thanks stovicek\@kmdec.fjfi.cvut.cz \endthanks
\thanks vittot\@cpt.univ-mrs.fr \endthanks

\abstract
We consider a perturbed Floquet Hamiltonian $-i\partial_t + H + \beta V(\omega t)$ 
in the Hilbert space $L^2([0,T],\Cal{H}$$,dt)$. Here $H$ is a self-adjoint operator 
in $\Cal{H}$ with a discrete spectrum obeying a growing gap condition, $V(t)$ is a 
symmetric bounded operator in $\Cal{H}$ depending on $t$ $2\pi$-periodically, 
$\omega = 2\pi/T$ is a frequency and $\beta$ is a coupling constant. The spectrum 
$\operatorname{Spec}(-i\partial_t + H)$ of the unperturbed part is pure point and 
dense in $\Bbb{R}$ for almost every $\omega$. This fact excludes application of the 
regular perturbation theory. Nevertheless we show, for almost all $\omega$ and 
provided $V(t)$ is sufficiently smooth, that the perturbation theory still makes 
sense, however, with two modifications. First, the coupling constant is restricted 
to a set $I$ which need not be an interval but 0 is still a point of density of 
$I$. Second, the Rayleigh-Schrodinger series are asymptotic to the perturbed 
eigen-value and the perturbed eigen-vector.
\endabstract
\endtopmatter

\document

CPT-97/P.3559

November 18, 1997

ftp://cpt.univ-mrs.fr

http://www.cpt.univ-mrs.fr

\newpage

\bigpagebreak
\flushpar {\bf 1. Introduction}
\medpagebreak

The so called Floquet Hamiltonians were introduced by Howland
\cite{\HowlandA} and Yajima \cite{\Yajima} in order to study
time-dependent quantum systems described by an Hamilton operator $H(t)$
acting in a Hilbert space $\HH$. Already before this strictly
mathematical setting of the problem one could meet similar ideas in the
physical literature \cite{\Sambe}. In our paper we restrict ourselves to
$T$-periodic time-dependent Hamiltonians. In this case the Floquet
operator is formally written as $K=-i\partial_t+H(t)$ and it acts in the
Hilbert space $\KK=L^2([\,0,T\,],\HH,dt)$. Usually $H(t)$ is decomposed
into a sum of a time-independent part $H$ and a time-dependent
perturbation $\beta V(\omega t)$ where $\omega=2\pi/T$ and $\beta$ is a
parameter (coupling constant). The primary question to be answered is
that of the character of the spectrum of $K$ \cite{\EnssVeselic}. What
makes this task difficult is the fact that, in many interesting
situations, the spectrum of the Floquet Hamiltonian associated to the
unperturbed (time-independent) Hamiltonian $H$ is pure-point and dense
in $\BR$. Particularly this excludes application of the regular
perturbation theory due to Rellich \cite{\Rellich} and Kato
\cite{\Kato}. Let us mention a few landmarks (but definitely not all of
them) in the comparatively short history of the problem which have
motivated us to deal with this subject.

In the article \cite{\Bellissard} Bellissard introduced a technique to
study time-dependent Schr\"odinger equations which was inspired by the
method of the proof of the classical KAM theorem
\cite{\Kolmogorov, \Arnold, \Moser}.
He considered a model on the circle (in which $\HH=L^2(S^1)$ and
$H=-\Delta$ with periodic boundary conditions) and he looked for
sufficient conditions to get pure-point spectrum of the
associated Floquet Hamiltonian. The density of the unperturbed
spectrum leads to a small divisors problem which was mastered in this
paper, for appropriate diophantine frequencies $\omega$ and $V$'s small
enough, by a method similar to the original KAM algorithm. We note that
Bellissard considered a perturbation $V$ acting as a multiplication
operator
by a function analytic both in the time and in the spatial variable.

Soon after Combescure addressed in \cite{\Combescure} the same
question, with $H$ being the one-dimensional harmonic oscillator
and $V$ not necessarily
analytic. To cope with the lack of analyticity she has adapted the
Nash-Moser regularization trick
\cite{\Moser}. However she had to face a more severe problem:
the spectrum of $H$ did not satisfy a growing gap condition
(this is an important technical property
which was satisfied in the Bellissard's model). This is why she had
to restrict the class of admissible perturbations, particularly
excluding realistic local potentials.
Let us mention also the work \cite{\BleherJauslinLebowitz} devoted to
an interesting model with constant gaps in the spectrum of $H$ and
with an analytic perturbation $V$.

Later on, the first two authors of the present paper considered in
\cite{\DS1} the same question in a more abstract situation: $H$ is
discrete, simple, with a growing gap condition (see formula (2.1)),
acting in a separable Hilbert space ${\Cal H}$ and with $V$ being not
necessarily analytic. More precisely, one didn't require that the
matrix entries of $V$ in the eigen-basis of $-i\partial_t+H$ were
exponentially decaying. The paper was based on a combination of two
methods: the Nash-Moser trick and the adiabatic regularization due to
Howland \cite{\HowlandB}. The latter method makes it possible, roughly
speaking, to convert the regularity of $V$ in the time variable into a
regularity in the spatial variable. For further development of this
procedure the reader can consult \cite{\Nenciu, \Joye}.
We note that in the reference \cite{\HowlandB} Howland proposed
another way to prove the pure-point character of a spectrum which
was based on a "randomization" of the original operator but he
failed to extend this results to the case when $H$ was a 
Schr\"odinger operator.

Two main characteristics are common to all the above works.
First, the results are global in the sense that they describe the
character of the full spectrum. Second, all these approaches are based
on the accelerated convergence method which is of iterative nature.
In fact, this method is an adaptation of a procedure used in
the celebrated KAM result concerning perturbations of classical
integrable systems. The present paper has another goal and an
essentially different method was necessary to reach it.
Here we concentrate on one single eigen-value. More precisely,
for operators of the same type as in \cite{\DS1}
we shall answer affirmatively the question: {\it Is it possible
to show that one single unperturbed eigen-value gives rise to an
eigen-value of the perturbed operator?}. We shall do it using a direct
method, this is to say, by showing directly that the standard 
eigen-value equation has a solution at least for appropriate values of 
the coupling constant $\beta$.

In our approach the eigen-vector is written in a form of an infinite
series and to verify its convergence we again have to cope with the
small divisors problem (see equation (3.3)). However we don't use any
kind of iterative methods and instead we rearrange partially the series
and estimate its summands directly. This compensation method was probably
more explicit in our previous paper \cite{\DSV1}. This article was
inspired by the pioneering work of Eliasson \cite{\Eliasson}
(see also an earlier paper by Siegel \cite{\Siegel}) and its
purpose was to check some basic ideas on an explicit example. Here
we treat the general case but we borrow from \cite{\DSV1} some
intermediate results, particularly this concerns Proposition 3.1 below.
Apart of the rearrangement of the series we use another crucial
technical trick. This is a sort of a reduction procedure
based on the observation that the eigen-values of the unperturbed
Floquet Hamiltonian which may be suspected to contribute by small
denominators are rather rare (see Sections 5 and 6). We note that this
idea, in a bit heuristic version, already appeared in the physical
literature \cite{\OGC}.

The paper is organized as follows. The main result (Theorem 2.1) is
formulated in the very beginning, i.e., in Section 2. The proof is split
into several steps which are carried out in the remainder of the paper,
i.e., in Sections 3-8. In fact, already after reading Section 3 one can
guess about the structure of the proof. Its summary is given at the end
of Section 8. The paper contains three appendices. In Appendix A we
present, for the sake of completeness, a proof of the fact that the
spectrum of the unperturbed Floquet Hamiltonian is dense in $\BR$ for
almost all frequencies. In Appendix B we construct an example of a
perturbation for which the formal solution of the equation on
eigen-values (so called Rayleigh-Schr\"odinger series) doesn't exist.
Appendix C contains a summary of the results about Lipschitz functions
that we need for our approach.

\bigpagebreak
\flushpar {\bf 2. The problem and the result}
\medpagebreak

Our goal is to study a perturbed Floquet Hamiltonian $K+\beta V$ acting 
in 
$$ 
\KK:=L^2([0,T],dt)\otimes\HH 
$$ 
where $\HH$ is a given separable Hilbert space, 
$$ 
K:=-i\pr_t\otimes1+1\otimes H 
$$ 
is the unperturbed (time-independent) part 
and $\beta$ is a coupling constant. We assume that $V(t)$ 
is a given $2\pi$-periodic sufficiently smooth function with values in 
the space of bounded operators $\BB(\HH)$, and $V(t)$ is symmetric for
all $t$. The perturbation $V$ is naturally induced by the
$T$-periodic function $V(\w t)$, with $\w:=2\pi/T$ being the frequency, 
and it is, of course, bounded and self-adjoint. We assume further that 
$H$ is a self-adjoint operator in $\HH$, its spectrum 
$$ 
\Spec(H)=\{E_k;\ k\in\BN\} 
$$ 
is discrete, simple and obeys the gap condition 
$$ 
\inf_{k\in\BN} {E_{k+1}-E_k\over (k+1)^{\alpha}}\ge C_E
\tag\tggap 
$$ 
where $C_E$ and $\al$ are strictly positive constants.

Here and everywhere in what follows we adopt the convention  
according to which 
$\BN$ stands for the set of natural numbers starting from 1 
whereas $\BZ_+$ includes also 0. 

As usual, we assume the periodic boundary conditions in time. 
The operator $K$ is self-adjoint and its spectrum equals
$$ 
\Spec(K)=\{F_n:=\w n_1+E_{n_2};\ n\in\BZ\times\BN\}.
$$ 
Denote by $f_n$, $n\in \BZN$, the corresponding normalized eigen-vectors 
and by $P_n$ the orthogonal projector onto $\BC f_n$. With the help of 
this eigen-basis we identify the Hilbert space $\KK$ with 
$l^2(\BZN)$ and all relevant operators with their matrices. Particularly 
the perturbation $V$ is represented by the matrix $(V_{mn})$, 
$$ 
\align 
V_{mn}  &= {1\over T}\int_0^T\langle e_{m_2},V(\w t)e_{n_2}\rangle_\HH
\,\exp\bigl(i\w(n_1-m_1)t\bigr)\, dt \\ 
&= {1\over 2\pi}\int_0^{2\pi}\langle e_{m_2},V(t)e_{n_2}\rangle_\HH
\,\exp\bigl(i(n_1-m_1)t\bigr)\, dt\, , \tag\tgV 
\endalign$$ 
where $\{e_k;\ k\in\BN\}$ denotes the orthonormal eigen-basis of $H$. 

Note that the matrix entries of $V$ don't depend on $\w$ and so the 
frequency occurs only in the eigen-values of $K$. The problem depends on
two parameters -- $\beta$ and also the period $T$. However, in the very
beginning, we shall fix $\w$ so that a diophantine condition
(cf. (\tgdiow)) is satisfied.
Afterwards we don't move the value $\w$ anymore and study the dependence 
only on the coupling constant. 

We have just presented all the incoming data. Let us now formulate the
problem. We fix once for all an index $\eta\in\BZN$ and write
$$
P:=P_\eta\quad\text{and}\quad Q:=1-P.
$$
Similarly, we redenote $f:=f_\eta$ and $F:=F_\eta$; so $Kf=Ff$ and
$Pf=f$, $Qf=0$, with $\|f\|=1$. We ask whether the operator $K+\beta V$
possesses also an eigen-value $F(\beta)$ which could be regarded as
being inherited from the eigen-value $F$ of $K$. The regular
perturbation theory due to Rellich \cite{\Rellich} and Kato \cite{\Kato}
in no way provides an answer to this question since the set
$\Spec(K)=\w\BZ+\Spec(H)$
is dense in $\BR$ for almost all $\w>0$. This property of the spectrum is
quite familiar, 
nonetheless we present an elementary proof in Appendix A. Recall
that the basic assumption for the regular theory to go through is that
the eigen-value $F$ is isolated. Also because of the density of the
spectrum, it makes little sense trying to relate, for a single value of
the parameter $\beta$, an eigen-value
$F(\beta)$ of $K+\beta V$ to the distinguished eigen-value $F$ of $K$.
But we shall show that it is reasonable to relate to $F$
a whole function $F(\beta)$, for $\bt$ running over some domain in the
vicinity of zero.

In our case, $F$ can be an accumulation point of $\Spec(K)$. On the
other hand, $F$ is a simple eigen-value for a generic $\w$ and so
the operator $K-F$ is injective on the subspace $\Ran(Q)$. In
fact, practically all subsequent manipulations will be concerned with
this subspace while the vector $f$ plays a role of a "source".
This is reflected in the notation; for an operator $X$ in $\KK$ we
denote by $\hat X$ its block corresponding to the subspace $\Ran(Q)$:
$$
\hat X:=QXQ\quad\text{ as an operator in }\Ran(Q). 
\tag\tghat
$$
Then $(\hK-F)^{-1}$ is a self-adjoint possibly unbounded operator.

There are more distinctions when comparing with the regular case. We
will discuss this point in a bit more detail in Section 3. Here we
recall that, according to the Rellich-Kato theorem -- the basic
result of the regular perturbation theory, if the eigen-value $F$
was simple and isolated then $F(\beta)$ would be an analytic function
on a neighbourhood of the origin. The same remark applies to the
eigen-vector $f(\beta)$ provided a convenient normalizing condition has
been imposed making it unambiguous. For example, a normalization
frequent in the physical literature \cite{\Messiah} requires that
$$
\langle f, f(\beta)\rangle =1\Longleftrightarrow
\fb-f\in\Ran(Q)
\tag{\tgnorm}
$$
is valid for all $\beta$ from the corresponding domain. The analytic
functions
$$
\align
F(\beta) &= F+\bt \la_1 +\bt^2\la_2 +\dots, \\
f(\bt) &= f+\bt g_1 +\bt^2 g_2 +\dots ,
\endalign
$$
are known as the Rayleigh-Schr\"odinger (RS) series, with the
coefficients $\la_j\in\BR$ and $\ g_j\in\Ran(Q)$ expressed explicitly
\cite{\Kato, \ReedSimon}. More details are given in Section 4. Here
we recall only that
$$
\la_1=\langle f,Vf\rangle,\ \la_2=-\langle QVf,(\hK-F)^{-1}QVf\rangle .
\tag{\tglam}
$$

On the contrary, this seems to be an intrinsic feature for the problems 
with dense point spectrum that the common domain $I$ for the functions 
$F(\bt)$ and $f(\bt)$ cannot be chosen as an interval. Because of the 
resonance effects it possesses numerous "holes". Nevertheless 0 can be
a point of density of $I$. Furthermore, the relation
of the RS series to the functions $F(\bt)$ and $f(\bt)$ is not 
straightforward. A priori it is even not clear whether the coefficients 
$\la_j$ and $g_j$ are well defined. For example, the existence of 
$\la_2$ in (\tglam) is guaranteed by the condition 
$QVf\in\Dom((\hK-F)^{-1})$ which is not obvious at all. Fortunately it 
turns 
out that the coefficients do exist, up to some order, provided $\Vt$ is 
sufficiently smooth. Then the RS series don't determine  
$F(\bt)$ and $f(\bt)$ directly but instead they describe the asymptotic 
behaviour of these functions as $\bt\to 0$.  

Now we are ready to formulate the result. Here $|X|$ stands for the
Lebesgue measure of a measurable set $X$.

\proclaim{Theorem 2.1}
Suppose that a self-adjoint operator $H$ with a discrete spectrum
obeys the gap condition (\tggap) and a
symmetric operator-valued function $\Vt\in C^r$ in the strong sense,
with $r\ge2$ and $r>16/\al$. Then there exists a set
$\Omega\subset\,]0,+\infty[$
of full Lebesgue measure such that, for all $\w\in\Omega$ and
any $\eta\in\BZN$ fixed, the Rayleigh-Schr\"odinger coefficients
$\la_j\in\BR$ and $g_j\in\Ran(Q)$, $1\le j\le\ell$, are well defined,
with $\ell$ being the greatest integer which fulfills
$$
\ell< {r\al\over 4}-2.
\tag\tgel
$$

If, moreover, the second coefficient $\la_2\not=0$ (as given in (\tglam))
then there exist a real function $F(\bt)$ and a $\KK$-valued function
$f(\bt)$ defined on a common domain $I$ and having the properties:

\noindent
(1)\phantom{\hskip 3pt}
$\fb\in\Dom(K)$, $\langle f,\fb\rangle=1$, 
and $(K+\bt V)\fb=\Fb\fb$ for all $\bt\in I$,\newline
(2)\phantom{\hskip 3pt}
$\lim_{\delta\downarrow0}|I\cap[\,-\delta,\delta\,]|/2\delta
=1$,\newline
(3)\phantom{\hskip 3pt}
$\Fb=F+\bt\la_1+\dots+\bt^\ell\la_\ell+O(|\bt|^{\ell+1})$,
\newline \phantom{(3)\hskip 3pt}
$\fb=f+\bt g_1+\dots+\bt^\ell g_\ell+O(|\bt|^{\ell+1})$.
\endproclaim

From the construction of the set $\Omega$ (c.f. (\tgOmega) and
Proposition 3.1) it is evident that the eigen-value $F$ of $K$ is simple
for all $\w\in\Omega$. Furthermore,
let us note that if $\Vt\in C^\infty$ then the coefficients $\la_j$
and $g_j$ exist for all $j\in\BN$ and the property (3) means that
the functions $\Fb$ and $\fb$ have asymptotic expansions at $\bt=0$
coinciding with the RS series.

We conclude this section by a brief comparison of this theorem with some
previous results. This concerns, first of all, the mutual role of
the two parameters $\omega$ and $\beta$. A notable approach to the
spectral problem of the operator $K+\beta V$ goes back to Bellissard
\cite{\Bellissard}
(see also \cite{\Combescure}, \cite{\DS1}). Also in this case, the
spectrum of the unperturbed Hamiltonian $H$ was supposed to obey
the same type of gap condition (\tggap). Under some
smoothness assumptions on $V(t)$, one is able to show that, for each
sufficiently small $\beta$, there exists a set of ``non-resonant''
frequencies $\Omega(\beta)$ such that the Lebesgue measure
of the complement of $\Omega(\bt)$ is reasonably small and the 
operator $K+\beta V$ is pure point for each
$\omega\in\Omega(\bt)$. The dependence
of $\Omega(\beta)$ on $\beta$ is to be emphasized. On the contrary,
the above theorem focuses only on one distinguished eigen-value.
But in this case one can choose the set $\Omega$ independently of
$\beta$ so that it covers almost all frequencies $\omega>0$ in the
Lebesgue sense. The basic problem now is to construct a convenient
domain $I$ for the coupling constant $\beta$, with $\w\in\Omega$ being
fixed. Naturally $I$ depends on the choice of the unperturbed
eigen-value.

We split the proof of Theorem 2.1 into several steps, each of them treated 
in one of the subsequent sections. A summary of all the steps is given at
the end of Section 8. 

\bigpagebreak
\flushpar {\bf 3. Projection method, comparison with the regular case}
\medpagebreak

We start the proof of Theorem 2.1 from the perturbed equation on
eigen-values,
$$
(K+\bt V)(f+g)=(F+\la)(f+g),
\tag 3.1
$$
with $\la\in\BR$ and, according to the normalization (\tgnorm),
$g\in\Ran(Q)$. Applying to (3.1) the complementary projectors $P$ and 
$Q$ (commuting with $K$) we obtain an equivalent set of equations (recall 
(\tghat)) 
$$
\gather
\la = \bt\langle Vf,f\rangle+\bt\langle Vf,g\rangle, \tag\tgsetIa\\
(\hK+\bt\hat V-F-\la)g = -\bt QVf\,. \tag\tgsetIb
\endgather
$$
For a while we shall consider $\la$ as another auxiliary parameter and
we  will try to solve the equation (\tgsetIb), referred to as the
eigen-vector equation from now on. Its solution is a vector-valued 
function 
$g=g(\bt,\la)$ depending on both parameters $\bt$ and $\la$, and
taking values in $\Ran(Q)$. Plugging $g(\bt,\la)$ into the equality  
(\tgsetIa) we get an implicit equation $\la=G(\bt,\la)$ from which one 
should extract a function $\la=\la(\bt)$. Then 
$$
\Fb=F+\la(\bt)\quad\text{and}\quad \fb=f+g(\bt,\la(\bt))
\tag 3.4
$$
will be the sought solution to our problem. This projection method was 
rediscovered many times  in the past and bears various names: 
Brillouin-Wigner, Feshbach, Grushin, Schur, $\dots$.

Naturally this procedure can be applied to the regular case as well and 
one 
can rederive this way the Rellich-Kato theorem. In order to emphasize the 
difference between the regular and non-regular cases we sketch below the 
basic steps. But before doing it let us introduce some more notation used 
throughout the paper. Set 
$$
\align
\Gamma_0 &:=(\hK-F)^{-1} , \tag\tgGO \\
\Ga_\la &:=(\hK-F-\la)^{-1}=(1-\la\Ga_0)^{-1}\Ga_0 . \tag\tgGla \\
\endalign
$$
Thus $\GO$ is a self-adjoint operator acting in $\Ran(Q)$ provided $F$ is 
a simple eigen-value of $K$. The same holds true for
$\Ga_\la$ if $\la\not\in\Spec(\hK-F)$. 

The regular case is characterized by the condition
$$
\text{dist}(F,\Spec(K)\setminus\{F\})=:d>0 .
\tag 3.7
$$
Hence the operator $\GO$ is even bounded and $\Vert\Gamma_0\Vert=d^{-1}$. 
Moreover, $\Ga_\la$ is bounded as well and depends analytically on $\la$ 
in the domain $\vert\lambda\vert<d$. However $K$ itself need not be
bounded and one can even consider a more general situation with $V$
being relatively bounded with respect to $K$. This assumption implies 
that 
$\Vert\Gamma_0\hat V\Vert=\Vert\hat V\Gamma_0\Vert<\infty$ and  it is 
sufficient to ensure that the operator 
$1+\beta\Gamma_\lambda\hat V$ is invertible provided the parameters 
$\bt$ and $\la$ belong to the domain   
$$
d\,\Vert\hat V\Gamma_0\Vert\,\vert\beta\vert+\vert\lambda\vert<d \,.
\tag 3.8
$$
Consequently, there exists a unique solution to (\tgsetIb) given by 
$$
g(\beta,\lambda)=-\beta(1+\beta\Gamma_\lambda\hat V)^{-1} 
\Gamma_\lambda QVf \,.
$$
Obviously, the function $g\bl$ is analytic in the domain (3.8) and 
its values belong to 
$\Dom(\hK-F-\la)\subset\Dom(K)$. The equality (\tgsetIa) then leads to 
the implicit equation
$$
\align 
& \la=G\bl,\quad\text{with}\\ 
& G(\beta,\lambda)=\beta\langle Vf,f\rangle-\beta^2 
\langle QVf,(1+\beta\Gamma_\lambda\hat V)^{-1}QVf\rangle \,.
\endalign
$$
Since $G(\beta,\lambda)$ is analytic and 
$$
(\lambda-G(\beta,\lambda))\vert_{(\beta,\lambda)=(0,0)}=0,\quad 
\partial_\lambda
(\lambda-G(\beta,\lambda))\vert_{(\beta,\lambda)=(0,0)}=1 \,,
$$
the implicit mapping theorem tells us that there exists 
a unique analytic function $\lambda=\lambda(\beta)$ defined on a
neighbourhood of the origin and such that $\lambda(0)=0$, 
$\lambda(\beta)=G(\beta,\lambda(\beta))$. In accordance with (3.4) we
get  both the perturbed eigen-value $\Fb$ and the eigen-vector $\fb$
as uniquely determined analytic functions.

Let us return to our problem with dense point spectrum and with $V$
being a bounded perturbation. Violation of the condition (3.7) means
exactly that the operator $\GO$ is unbounded. We shall need another
but weaker condition in order to be still able to cope with the
equation (\tgsetIb). Diophantine estimates are the standard tool
used widely in this situation. Let us first 
introduce the relevant exponents. The integer $\ell$, as specified in 
Theorem 2.1 (cf. (\tgel)), obeys 
$$
\ell\ge2\quad\text{and}\quad 4\ell+8<r\al.
$$
Hence one can find reals $\tau>4$ and $\sigma>1$ such that 
$$
\tau(\ell+2)\le r\al\quad\and\quad 2\sg+2<\tau.
\tag\tgexpo
$$
Next we define the set of non-resonant frequencies,
$$
\Omega_\eta:=\{\w>0;\ \inf_{n\in\BZN,\ n\not=\eta}
n_2^{\,\sg}|F_n-F|>0\}. 
\tag\tgOmega
$$
A simple adaptation of the proof of Lemma 4 in \cite{\DSV1} shows that
if $\sg>1$ then $\Omega_\eta\subset\,]0,+\infty[$ is of full Lebesgue
measure. It is clear that a non-resonant frequency can be even chosen
for all indices $\eta$ simultaneously.

\proclaim{Proposition 3.1} 
Suppose that $\sg>1$. Then almost all $\w>0$ belong to
$$
\Omega:=\bigcap_{m\in\BZN}\Omega_m \,.
$$
\endproclaim 

We fix, for the rest of the paper, a non-resonant frequency 
$\w\in\Omega$. 
Then eigen-values of $\hK-F$ fulfill the diophantine estimate  
$$
|F_n-F|\ge\gamma\,n_2^{\,-\sg}\quad\fora\ n\in\BZN,\
n\not=\eta,
\tag\tgdiow
$$
with some constant $\ga>0$. In addition, 
the property (\tgdiow) guarantees that $F$ is 
a simple eigen-value of $K$. We shall write 
$$
\psi(k):=\ga\, k^{-\sg},\quad \tpsi(k):={\ga\over 2}\,k^{-\tau}\,.
\tag\tgpsi
$$

We would like to warn the reader that, in order to avoid introducing 
additional symbols,  
the restrictions (\tgexpo) on $\tau$ 
will be applied in the subsequent procedure only at those places where 
they have some consequences, otherwise $\tau$ can be any real number. 
Similarly, $\ell$ can be any non-negative integer if not specified 
otherwise. 

Let us finish shortly the comparison of the regular and non-regular cases 
by indicating some forthcoming steps. The discrete function $\tpsi(k)$ 
given in (\tgpsi) will be used later, in Section 6, in another 
diophantine 
estimate involving the parameters $\bt$ and $\la$ and defining a closed 
set $\DD\subset\BR^2$. We shall be able to solve the eigen-vector 
equation (\tgsetIb) provided $\bl\in\DD$ getting
this way a vector-valued function 
$g\bl$. Consequently the function $G\bl:=\bt\,\langle Vf,g\bl\rangle$ 
is defined only on the set $\DD$, too, but fortunately one can show that 
$G$ belongs to the Lipschitz class $\Lip(\ell+1,\DD)$, with $\ell$ 
specified in Theorem 2.1. This enables one to apply the Whitney 
extension theorem in order to extend $G$ from $\DD$ to $\BR^2$. 
Making the standard simplifying assumption that $\langle Vf,f\rangle=0$ 
one again arrives at the implicit equation $\la=\tG\bl$, with the 
extended right 
hand side. The implicit mapping theorem guarantees the existence of a 
solution $\la=\tla(\bt)$. However one has to restrict the function $\tla$ 
to the set $I$ determined by the condition $(\bt,\tla(\bt))\in\DD$. Thus 
the resulting function $\la(\bt)$ is not defined on an interval but, 
on the other hand, one can verify that its domain $I$ is still reasonably 
dense at the origin. 

\bigpagebreak
\flushpar {\bf 4. Perturbation series}
\medpagebreak

In this section we summarize a few basic facts about the RS series, 
particularly we recall the explicit expressions for coefficients in a
form relying on some combinatorial notions. Basically we adopt the
physical
point of view according to which one seeks the eigen-vector $\fb$ 
normalized by $\langle f,\fb\rangle=1$ \cite{\Messiah}. In a more 
mathematically oriented approach one prefers to treat the orthogonal 
projector $P(\bt)$ onto the 1-dimensional subspace $\BC\fb$ rather than 
the 
vector $\fb$ itself. Then the corresponding formulas take an optically 
different form \cite{\Kato}. But, of course, our choice is only a matter 
of 
taste and convenience as the both approaches are obviously equivalent; 
for example 
$$
\fb={1\over \langle f,P(\bt)f\rangle}\,P(\bt)f\,.
$$
On the other hand, the eigen-value $\Fb$ is unambiguous and the result 
must be the same in any case. This point has been discussed shortly in 
\cite{\DSV1}. 

We are forced to use a bit more general setting since the functions $\Fb$ 
and $\fb$ need not be analytic and instead they are characterized by
their asymptotics. However this doesn't cause a serious complication.

\proclaim{Lemma 4.1}
Suppose that 0 is an accumulation point of a closed set $I\subset\BR$, 
$\ell\in\BN$, and we are given a real function $\Fb$ and a $\KK$-valued 
function $\fb$, both defined on $I$ and having asymptotics at $\bt=0$: 
$$
\align
\Fb &= F+\bt\la_1+\dots+\bt^\ell\la_\ell+O(|\bt|^{\ell+1}),
\tag 4.1 \\
\fb &= f+\bt g_1+\dots+\bt^\ell g_\ell+O(|\bt|^{\ell+1}).
\tag 4.2 \\
\endalign 
$$
Suppose, moreover, that for all $\bt\in I$, $\fb\in\Dom(K)$ and
$$
(K+\bt V)\fb=\Fb\fb\,.
\tag 4.3 
$$
Then $f,g_1.\dots,g_\ell\in\Dom(K)$ and 
$$
K\fb = Kf+\bt\,Kg_1+\dots+\bt^\ell\,Kg_\ell+O(|\bt|^{\ell+1}). 
\tag 4.4 
$$
\endproclaim
\demo{Proof}
The function $K\fb$ has an asymptotic as well since 
$$
K \fb=-\bt\,V\fb+\Fb\fb=u_0+\bt u_1+\dots+\bt^\ell u_\ell+
O(|\bt|^{\ell+1}).
$$
Redenote temporarily $f$ as $g_0$. Proceeding by induction in $j$ we shall 
show that $g_j\in\Dom(K)$ and $Kg_j=u_j$, $j=0,1,\dots,\ell$. This is 
obvious for $j=0$ as $g_0=f(0)$ and $Kg_0=Kf(0)=u_0$. Suppose that $j\ge1$ 
and set temporarily, for $\bt\not=0$, 
$$
h_j(\bt):=\bt^{-j}\left(\fb-g_0-\bt\,g_1-\dots-\bt^{j-1}\,g_{j-1}\right).
$$
Then $h_j(\bt)\to g_j$ and, by the induction hypothesis, 
$h_j(\bt)\in\Dom(K)$ and $Kh_j(\bt)\to u_j$, as $\bt\to0$. But $K$ is 
closed and so $g_j\in\Dom(K)$ and $Kg_j=u_j$. \qed
\enddemo 

From the existence of the asymptotics (4.1), (4.2) and (4.4) follows 
immediately that the corresponding coefficients on the both sides of 
(4.3) coincide up to the order $\ell$. This leads to the system of 
equations ($g_0\equiv f$) 
$$
(K-F)g_M=-Vg_{M-1}+\sum_{j=1}^{M-1}\la_j\,g_{M-j}+\la_M\,f,\quad 
1\le M\le\ell.
\tag\tgeqI
$$
If $\fb$ obeys the normalization (\tgnorm), and so $g_j\in\Ran(Q)$ for 
$j\ge1$, one can again separate the parts belonging to $\Ran(P)$ and 
$\Ran(Q)$ getting this way 
$$
\aligned
& (\hK-F)g_M=-\hV g_{M-1}+\sum_{j=1}^{M-1}\la_j\,g_{M-j}
\quad\text{where}\\ 
& \la_M=\langle QVf,g_{M-1}\rangle,\ M=1,\dots,\ell,\\
\endaligned
\tag\tgeqII
$$
(for $M=1$, $\hV g_{M-1}$ should be replaced by $QVf$). We still assume 
that 
$(\hK-F)^{-1}=\GO$ exists. Clearly one can calculate, successively and 
unambiguously, the vectors $g_1,\dots,g_\ell$, and consequently the 
numbers $\la_1,\dots,\la_\ell$ as well provided one can show that 
$g_1,\dots,g_{\ell-1}$ and $QVf,\hV g_1,\dots,\hV g_{\ell-1}$ belong to 
$\Ran(\GO)$. In this case we can rewrite (\tgeqII) in the form 
$$
g_M=-\GO\hV g_{M-1}+\sum_{j=1}^{M-1}\la_j\,\GO g_{M-j},\quad 
M=1,\dots,\ell. 
\tag\tgeqIII 
$$
One deduces readily from (\tgeqIII) that $g_M$ is a linear combination of 
the vectors 
$$
\GO^{\,s_1}\hV\dots\hV\GO^{\,s_p}QVf\quad\text{with}\ 1\le p\le M,\ 
(s_1,\dots,s_p)\in\BN^p,\ \and\ \sum s_i\le M.
\tag\tgvectors
$$
Hence the existence of vectors (\tgvectors), for $M=1,\dots,\ell$, 
represents a sufficient condition for the system (\tgeqII) to have a 
unique solution. 

Before approaching the explicit expressions let us recall a bit of 
combinatorics. The set of rooted $N$-trees $\TT(N)\subset\BZ_+^N$ is 
characterized by the condition ($|\nu|:=\nu_1+\dots+\nu_N$): 
$$
\align
& \nu=(\nu_1,\dots,\nu_N)\in\TT(N)\Longleftrightarrow \\
& \nu_k + \dots + \nu_N\le N-k\ \for\  
2\le k\le N,\ \and\ |\nu|=N-1.
\endalign
$$
Obviously $\nu_N=0$, and if $N\ge2$ then $\nu_1\ge1$. It is also quite 
easy to verify a composition rule for two trees, namely 
$$
\nu'\in\TT(N'),\ \nu''\in\TT(N'')\Longrightarrow 
\nu=(\nu',\nu'')+(1,0,\dots,0)\in\TT(N'+N'').
$$
As stated in the following lemma this procedure is invertible. We don't 
recall the proof. 

\proclaim{Lemma 4.2}  
Suppose that $\nu\in\TT(N)$ and $N\ge2$. Then there exists a unique 
decomposition $\nu=(\nu',\nu'')+(1,0,\dots,0)$ where 
$\nu'\in\TT(N')$, $\nu''\in\TT(N'')$ and $N'+N''=N$.
\endproclaim 

Now we are ready to describe the solution to the system (\tgeqII).

\proclaim{Proposition 4.3} 
Suppose that the vectors 
$\GO^{\,s_1}\hV\dots\GO^{\,s_{p-1}}\hV\GO^{\,s_p}QVf$ 
are well defined for all $p\in\BN$, $1\le p\le \ell$, and all
$p$-tuples $(s_1,\dots,s_p)\in\BN^p$ such that $\sum s_i\le \ell$. Then 
there exists a unique $\ell$-tuple $g_1,\dots,g_\ell$ solving the system 
of equations (\tgeqII). 

Suppose, in addition, that $\langle Vf,f\rangle=0$. Then the solution is 
given by the formula ($1\le M\le\ell$)
$$
g_M=\sum_{N\in\BN}\ \sum_{\nu\in\TT(N)}\ \sum_{k(1),\dots,k(N)\in\BN}\ 
\sum_{\mu(1)_\in\BN^{k(1)},\dots,\mu(N)_\in\BN^{k(N)} } 
\gG_M(N,\nu,k(j),\mu(j))
\tag\tggMa
$$
where the range of summation is restricted by the conditions
$$
k(1)+\dots+k(N)+N=M+1,\ |\mu(j)|=k(j)+\nu_j\quad\for\  
1\le j\le N\,,
\tag\tggMb 
$$
and
$$
\aligned
\gG_M(N,\nu,k(j),\mu(j)) :=& (-1)^{M+N+1}\prod_{j=2}^N
\langle Vf,\, \GO^{\mu(j)_1}\hV\GO^{\mu(j)_2}\dots
\hV\GO^{\mu(j)_{k(j)} }Vf\rangle \\
& \times \GO^{\mu(1)_1}\hV\GO^{\mu(1)_2}\dots
\hV\GO^{\mu(1)_{k(1)} }Vf \,.
\endaligned
\tag\tggMc
$$
The numbers $\la_1,\dots,\la_\ell$ are given correspondingly by 
$\la_1=\langle Vf,f\rangle=0$ and, for $2\le M\le\ell$,
$$
\aligned
\la_M &= \langle Vf,g_{M-1}\rangle\\ 
&=\sum_{N\in\BN}\ \sum_{\nu\in\TT(N)}\ \sum_{k(1),\dots,k(N)\in\BN}\ 
\sum_{\mu(1)_\in\BN^{k(1)},\dots,\mu(N)_\in\BN^{k(N)} } 
\gL_M(N,\nu,k(j),\mu(j))
\endaligned
\tag\tglaMa
$$ 
where the range of summation is restricted by the conditions
$$
k(1)+\dots+k(N)+N=M,\ |\mu(j)|=k(j)+\nu_j\quad\for\  
1\le j\le N\,,
\tag\tglaMb 
$$
and
$$
\gL_M(N,\nu,k(j),\mu(j)) := (-1)^{M+N}\prod_{j=1}^N
\langle Vf,\, \GO^{\mu(j)_1}\hV\GO^{\mu(j)_2}\dots
\hV\GO^{\mu(j)_{k(j)} }Vf\rangle \,.
\tag\tglaMc 
$$
\endproclaim
\demo{Proof}
The first part of the proposition has been discussed above. Let us show 
that the vectors $g_M$ given in (\tggM) obey the relation (\tgeqIII). 
This is easy to check for $M=1$. Then necessarily $N=1$ and so, 
as $\TT(1)=\{(0)\}$, the formula (\tggM) gives the correct answer 
$g_1=-\GO Vf$. Suppose that $M\ge2$. Observe that the assumption 
$\langle Vf,f\rangle=0$ implies that $\la_1=0$ and so the summation index 
on the RHS of (\tgeqIII) starts from the value $j=2$. Moreover, 
$Vf\in\Ran(Q)$. The verification is 
based on the following two equalities. First, 
$$
-\GO\hV\gG_{M-1}(N',\nu',k(j)',\mu(j)')=\gG_M(N,\nu,k(j),\mu(j))
\tag 4.11a 
$$
where
$$
\aligned
& N=N',\ \nu=\nu'\in\TT(N),\ k(1)=k(1)'+1,\ k(2)=k(2)',\dots,
k(N)=k(N)',\\ 
& \mu(1)=(1,\mu(1)'),\ \mu(2)=\mu(2)',\dots,\mu(N)=\mu(N)'.\\
\endaligned
\tag 4.11b 
$$
Second, if $1\le M',$ $2\le M''$ and $M=M'+M''$ then 
$$
\gL_{M''}(N'',\nu'',k(j)'',\mu(j)'')
\gG_{M'}(N',\nu',k(j)',\mu(j)')=\gG_M(N,\nu,k(j),\mu(j))
\tag 4.12a
$$
where
$$
\aligned 
& N=N'+N'',\ \nu=(\nu',\nu'')+(1,0,\dots,0)\in\TT(N), \\
& k(1)=k(1)',\dots,k(N')=k(N')',
k(N'+1)=k(1)'',\dots,k(N'+N'')=k(N'')'', \\
& \mu(1)=\mu(1)'+(1,0,\dots,0),\ 
\mu(2)=\mu(2)',\dots,\mu(N')=\mu(N')', \\
& \mu(N'+1)=\mu(1)'',\dots,\mu(N'+N'')=\mu(N'')''. 
\endaligned 
\tag 4.12b
$$

On the other hand, consider a summand $\gG_M(N,\nu,k(j),\mu(j))$. We 
distinguish two cases. If $\mu(1)_1=1$ then necessarily $k(1)\ge2$
and there exists a unique multiindex 
$(N',\nu',k(1)',\dots,k(N')',\mu(1)',\dots,\mu(N')')$ determining a 
summand $\gG_{M-1}$ such that (4.11) holds. If $\mu(1)_1\ge2$ then 
necessarily $N\ge2$ and, in virtue of Lemma 4.2, there exists a unique 
decomposition $\nu=(\nu',\nu'')+(1,0,\dots,0)$ where $\nu'\in\TT(N')$, 
$\nu''\in\TT(N'')$ and $N'+N''=N$. Set 
$$
M'=k(1)+\dots+k(N')+N'-1,\ M''=k(N'+1)+\dots+k(N)+N''.
$$
Observe that $N''\ge1$ implies $M''\ge2$. This way one obtains 
unambiguously  two multiindices $(N',\nu',k(j)',\mu(j)')$ and 
$(N'',\nu'',k(j)'',\mu(j)'')$ determining respectively summands 
$\gG_{M'}$ and $\gL_{M''}$ such that (4.12) holds. This completes the
verification. \qed 
\enddemo

\bigpagebreak
\flushpar {\bf 5. Set of critical indices, existence of the
Rayleigh-Schr\"odinger coefficients}
\medpagebreak

Let us continue the proof of Theorem 2.1. The arbitrarily small numbers
in $\Spec(\hK-F)$, so called small denominators, represent 
the principal difficulty we have encountered in the preceding discussion. 
This is why the operator $\GO=(\hK-F)^{-1}$ is not bounded and thus it 
is not a priori clear whether the assumptions of Proposition 4.3 are 
fulfilled and whether the RS coefficients exist at all. The second basic 
ingredient of our approach, apart of the projection method, is the 
observation that the indices suspected of enumerating small denominators 
are distributed rather rarely in the lattice $\BZN$. We introduce the 
set $\SS\subset\BZN\setminus\{\eta\}$ of ``critical'' indices by imposing 
the condition 
$$
n\in\SS\Longleftrightarrow
F_n-F\in\,\left]-{\w\over 2},{\w\over 2}\,\right] \,.
\tag\tgS
$$
Clearly, to each $n_2\in\BN$, $n_2\not=\eta_2$, there exists exactly one 
$n_1\in\BZ$ such that $(n_1,n_2)\in\SS$ and there is no such $n_1$ for 
$n_2=\eta_2$. In other words, the projection 
$\SS\to\BN\setminus\{\eta_2\}:$ $n\mapsto n_2$ is one-to-one. Roughly 
speaking, the indices from the set $\SS$ are situated closely to the 
curve $n_1=\eta_1+(E_{\eta_2}-E_{n_2})/\w$.

Now the gap condition (\tggap) can be employed to get more information 
about the set $\SS$. It is quite useful to observe that another inequality 
follows straightforwardly from (\tggap), namely 
$$
\left| E_j-E_k\right|\ge{C_E\over 1+\al}\,|j-k|\,
\max\{ j^\al,k^\al\},\quad \forall j,k\in\BN. 
\tag\tgineq
$$ 
Indeed, if $j>k$ then 
$$
\align 
E_j-E_k &= \sum_{p=k}^{j-1} (E_{p+1}-E_p)     
\ge C_E\,\int_k^j s^\al\, ds \\
&= {C_E\over 1+\al}\,(j^{1+\al}-k^{1+\al})    
\ge {C_E\over 1+\al}\,(j-k)\,j^\al \,. 
\endalign
$$ 
Using (\tgS) one derives that, for $m,n\in\SS$,
$$
\left| m_1-n_1\right|={1\over\w}\,\left|F_m-F_n-E_{m_2}+E_{n_2}\right| 
\ge {1\over\w}\,\left|E_{m_2}-E_{n_2}\right|-1\,.
\tag 5.3 
$$
A combination of (5.3) and (\tgineq) yields
$$
m,n\in\SS \Longrightarrow
1+\left|m_1-n_1\right|\ge {C_E\over\w(1+\al)}\, 
\max\{m_2^{\,\al},n_2^{\,\al}\}\,\left|m_2-n_2\right|\,.
\tag\tgdistI
$$
Similarly,
$$
m\in\SS\Longrightarrow 
1+\left|m_1-\eta_1\right|> {C_E\over\w(1+\al)}\, 
\max\{m_2^{\,\al},\eta_2^{\,\al}\}\,\left|m_2-\eta_2\right|\,.
\tag\tgdistII
$$

The set $\SS$ induces a splitting of the subspace $\Ran(Q)$ into the 
``singular'' and ``regular'' parts. This idea will be exploited more 
systematically in Section 6. Here we introduce the 
corresponding projectors, 
$$
P_S:=\sum_{n\in\SS} P_n,\quad P_R:=Q-P_S\,.
$$
Note that 
$$
\|(\hK-F)P_S\|\le{\w\over2},\quad \|\GO P_R\|\le{2\over\w}\,.
\tag\tgbound 
$$
Hence the restriction of $\GO$ to the subspace $\Ran(P_R)$ is quite 
harmless. 

Let us switch to the problem of RS coefficients. To show their existence, 
and also later in Section 6, we shall need an inequality with 
commutators. First we specify the underlying notions. Let $A$ be a 
closed, densely defined operator in $\KK$ and $X\in\BB(\KK)$. By saying 
that $\ad_A X$ is bounded we mean that: $\Dom(A)\subset\Dom(AX)$ and 
the operator $AX-XA$ is bounded on $\Dom(A)$, and so it can be unambiguously 
extended to an operator from $\BB(\KK)$ that we call $\ad_A X$. 
Particularly, $[\,A,X\,]=0$ is equivalent to: $\Dom(A)\subset\Dom(AX)$ 
and $AX=XA$ on $\Dom(A)$. One has the Leibniz rule in the following sense: 
if $X_1,X_2\in\BB(\KK)$ and both $\ad_A X_1,\ \ad_A X_2$ are bounded 
then so is $\ad_A(X_1X_2)$ and it holds 
$$
\ad_A(X_1X_2)=(\ad_A X_1)X_2+X_1(\ad_A X_2).
$$
More generally, saying that $\ad_A^{\ r}X$ is bounded, with $r\in\BZ_+$, 
means that: $\Dom(A^r)$ is dense in $\KK$, $\Dom(A^j)\subset\Dom(A^jX)$ 
for all $j,\ 0\le j\le r$, and the operator
$$
\sum_{j=0}^r \binom rj \,(-1)^j\,A^{r-j}XA^j\,,
\tag 5.7 
$$
clearly well defined on $\Dom(A^r)$, is bounded. We call the closure of 
(5.7) $\ad_A^{\,r}X$. The Leibniz rule can be generalized as usual. 

\proclaim{Lemma 5.1}
Suppose that we are given $p,r\in\BN$, a closed, densely defined operator 
$A$ and $X,B_1,\dots,B_{p-1}\in\BB(\KK)$ such that the operators 
$\ad_A^{\ j}X$ are bounded for all $j$, $1\le j\le r$, and 
$$
[\,A,B_1\,]=\dots=[\,A,B_{p-1}\,]=0.
$$
Then $\ad_A^{\ r}(XB_1X\dots B_{p-1}X)$ is bounded and its norm is 
estimated from above by 
$$
\aligned
\prod_{i=1}^{p-1}\|B_i\|\  
\sum\Sb \nu\in\BZ_+^r\\ \nu_1+2\nu_2+\dots+r\nu_r=r\endSb &
\ \frac{r!}{\prod\limits_{j=1}^r (j!)^{\nu_j}\,\nu_j!}\ 
p(p-1)\dots(p-|\nu|+1) \\ 
& \times\,\|X\|^{p-|\nu|}\,\prod_{j=1}^r\|\ad_A^{\ j}X\|^{\nu_j} \,.
\endaligned
\tag\tgadbound
$$
\endproclaim 
\demo{Proof} 
Let us recall a formula of differentiation of functions,
$$
\aligned
\pr_x^r\,h(x)^p =  
\sum\Sb \nu\in\BZ_+^r\\ \nu_1+2\nu_2+\dots+r\nu_r=r\endSb &
\ \frac{r!}{\prod\limits_{j=1}^r (j!)^{\nu_j}\,\nu_j!}\ 
p(p-1)\dots(p-|\nu|+1) \\ 
& \times\,h(x)^{p-|\nu|}\,\prod_{j=1}^r
\bigl(\pr_x^j\,h(x)\bigr)^{\nu_j} \,.
\endaligned
\tag 5.9
$$
In our case $\ad_A$ plays the role of differentiation. However, one 
cannot use the formula (5.9) directly since generally $\ad_A^{\ i}X$ 
and $\ad_A^{\ j}X$, for $i\not=j$, don't commute. Nevertheless we have, 
according to the generalized Leibniz rule, 
$$
\ad_A^{\ r}(XB_1X\dots B_{p-1}X)=
$$
$$
=\sum_{\mu\in\BZ_+^p,\ |\mu|=r}\binom r\mu \, 
\left(\ad_A^{\ \mu_1}X\right)B_1\left(\ad_A^{\ \mu_2}X\right)
\dots B_{p-1}\left(\ad_A^{\ \mu_p}X\right) .
\tag 5.10
$$
Estimating the norm of each summand in (5.10) by 
$$
\binom r\mu \,\prod_{i=1}^{p-1}\|B_i\|\, 
\prod_{j=1}^r\|\ad_A^{\ j}X\|^{\nu_j} 
$$
and grouping together the terms with the same powers $\mu_1,\dots,\mu_p$, 
up to a permutation, one arrives obviously at the same coefficients as 
in (5.9). \qed 
\enddemo

In the subsequent applications we substitute the time derivative for the 
operator $A$. Set $D:=(-i/\w)\,\pr_t\otimes1$; this is to say, when 
identifying $\KK\equiv l^2(\BZN)$, 
$$
Dh_n=n_1\,h_n,\quad\forall h=(h_n)\in\Dom(D)\subset\KK.
\tag\tgD
$$
It is clear that $D$ is reducible by the projectors $P$ and $Q$.
If $V(t)\in C^r$ then the operator-valued function $V^{(j)}(\w t)$, 
with $0\le j\le r$, induces naturally the bounded operator 
$\ad_D^{\ j}V\in\BB(\KK)$, and we have  
$$
\left(\ad_D^{\ j}V\right)_{mn}=\left(m_1-n_1\right)^j\,V_{mn}\,.
$$
This is a standard remark that the differentiability or, more generally, 
the boundedness of $\ad_D^{\ r}X$ induces a decay of matrix entries of 
an operator $X\in\BB(\KK)$. More precisely, if $X$ and $\ad_D^{\ r}X$ are 
bounded then 
$$
|X_{mn}|\le\max\{\|X\|,\ 2^r\,\|\ad_D^{\ r}X\|\}\,
\left(1+|m_1-n_1|\right)^{-r}\,.
\tag\tgdecay
$$
Particularly this applies to $V\in\BB(\KK)$. 

To proceed further we employ the diophantine estimate (\tgdiow). 

\proclaim{Lemma 5.2} 
Suppose that, in the strong sense, $V(t)\in C^r$, and $r\ge2$. Then for 
any $p$-tuple $(s_1,\dots,s_p)\in\BN^p$ and $q\in\BN$ such that 
$q\sg\le r\al$ it holds true that 
\roster 
\item"(i)" 
$\GO^q P_S(\hV \GO^{s_1}P_R\hV\dots\GO^{s_p}P_R\hV) P_S\GO^{-q} 
\in\BB(\Ran(Q))$, \newline 
\item"(ii)" 
$\GO^q P_S(\hV \GO^{s_1}P_R\hV\dots\GO^{s_p}P_R V)f$ is well defined.
\endroster

Both in (i) and (ii) the value $p=0$ is allowed and then the corresponding 
expressions read $\GO^q P_S\hV P_S\GO^{-q}$ and $\GO^q P_S Vf$, 
respectively. 
\endproclaim  
\demo{Proof}      
First we establish the inequality
$$
|(\hV \GO^{s_1}P_R\hV\dots\GO^{s_p}P_R\hV)_{mn}|\le 
\left({2\over\w}\right)^{\sum s_j}\, C_V\, (1+|m_1-n_1|)^{-r}
\tag 5.13  
$$ 
where $C_V\equiv C_V(p,r)$ is a constant. Indeed, according to
Lemma 5.1,
\newline
$\ad_{\hat D}^{\ r}(\hV \GO^{s_1}P_R\hV\dots\GO^{s_p}P_R\hV)$ is
bounded
for $[\,\hat D,\GO P_R\,]=0$ and $\ad_{\hat D}^{\ j}\hV$ are bounded, 
$1\le j\le r$. When applying the bound (\tgadbound) observe that 
$$
\prod_{j=1}^p \|\GO^{s_j} P_R\| \le 
\left({2\over\w}\right)^{\sum s_j}\,.
$$
Now it suffices to use (\tgdecay). 

We shall verify the item (i); the proof of (ii) is quite similar. Set 
temporarily 
$$
Y:=\GO^q P_S\hV \GO^{s_1}P_R\hV\dots\GO^{s_p}P_R\hV P_S\GO^{-q} \,.
$$
Suppose that $m,n\in\SS$. By the inequality (5.13) we have 
$$
|Y_{mn}|\le\const\,|F_m-F|^{-q}\,(1+|m_1-n_1|)^{-r}\,|F_n-F|^q\,.
\tag 5.14 
$$
The diagonal of $Y$ is bounded and so it suffices to estimate only the 
off-diagonal part. Combining (5.14) with (\tgdiow), (\tgbound) and 
(\tgdistI) we get ($m\not=n$) 
$$
|Y_{mn}|\le\const\,\left({\w(1+\al)\over C_E}\right)^r \,\ga^{-q}\, 
\left({\w\over2}\right)^q \,m_2^{\,q\sg-r\al}\,
|m_2-n_2|^{-r} \le \const'\,|m_2-n_2|^{-r} \,.
$$
Since $r\ge2$ we deduce that both 
$$
\sup_{m\in\SS}\sum_{n\in\SS} |Y_{mn}|\quad\and\quad
\sup_{n\in\SS}\sum_{m\in\SS} |Y_{mn}|
$$
are finite and, in accordance with the Schur-Holmgren criterion, the norm 
$\|Y\|$ is estimated from above by the maximal of these two numbers. \qed 
\enddemo 

As a straightforward consequence we get 

\proclaim{Lemma 5.3} 
Suppose that $V(t)\in C^r$, $r\ge2$. Then for any 
$p$-tuple $(s_1,\dots,s_p)\in\BN^p$ it holds true that 
$$
\sum_{j=1}^p s_j\le{r\al\over\sg}\Longrightarrow 
\GO^{s_1}\hV\dots\GO^{s_{p-1}}\hV\GO^{s_p}QVf\quad  
\text{is well defined}. 
$$
\endproclaim
\demo{Proof} 
Write, for each $j$, 
$$
\GO^{s_j}=\GO^{s_j}P_R+\GO^{s_j}P_S 
$$
and expand the resulting expression getting this way $2^p$ summands. 
Lemma 5.2 ad(i) can be used to move, in each summand, those powers 
$\GO^{s_{i_j}}$ which are accompanied by the projector $P_S$ from the left 
to the right. Thus the problem reduces finally to the existence of the 
vector 
$$ 
\GO^q P_S\hV \GO^{s_k}P_R\dots\GO^{s_{p-1}}P_R\hV\GO^{s_p}P_R Vf 
$$ 
where $1\le k\le p+1$ (by definition, the expression reads $\GO^qP_S Vf$ 
for $k=p+1$) and $q=\sum s_{i_j}$. By assumption, $q\le r\al/\sg$ and 
thus Lemma 5.2 ad(ii) proves the result. \qed 
\enddemo 

Combining Proposition 4.3 with Lemma 5.3 we get 

\proclaim{Proposition 5.4} 
Suppose that $\Vt\in C^r$, with $r\ge2$, and $\ell\in\BN$ obeys 
$\sg\ell\le r\al$. Then the Rayleigh-Schr\"odinger coefficients 
$\la_1,\dots,\la_\ell\in\BR$ and $g_1,\dots,g_\ell\in\Ran(Q)$ exist and 
represent the unique solution to  
the system of equations (\tgeqI) (or, equivalently, (\tgeqII)).
\endproclaim 

\demo{Remarks} 
(1) The existence of the RS coefficients is guaranteed by 
the differentiability 
of $\Vt$; the strong continuity is generally not sufficient. One can 
construct, for almost all $\w>0$, an operator-valued function $\Vt$ which 
is strongly continuous and such that already the coefficient $\la_2$ doesn't 
exist. This is the subject of Appendix B. 

\enddemo 
\demo\nofrills{} 
(2) For the choice of $\sg$ and $\tau$ specified in (\tgexpo) it holds 
clearly true that $\sg\ell<\tau(\ell+2)$ and hence the assumptions of 
Proposition 5.4 are fulfilled. So the first part of Theorem 2.1 has 
been proven. On the other 
hand, this comparison suggests that the assumption $r>16/\al$ of Theorem 2.1 
is very probably not optimal and could be improved. 
\enddemo 

\bigpagebreak
\flushpar {\bf 6. Solution of the eigen-vector equation}
\medpagebreak

In the sequel we adopt a standard simplification which doesn't imply any loss 
of generality. Namely, replacing $V$ by $V-V_{\eta\eta}$ means just the shift 
of the spectrum,
$$
\Spec(K+\bt(V-V_{\eta\eta}))=\Spec(K+\bt V)-\bt V_{\eta\eta}\,,
$$
while all eigen-vectors stay untouched. Also the assumptions of Theorem 2.1 
are not influenced by this replacement; particularly the coefficient $\la_2$ 
given in (\tglam) suffers no change (as $Qf=0$). So from now on we assume that 
$$
V_{\eta\eta}\equiv\langle Vf,f\rangle=0 \Longleftrightarrow 
Vf\in\Ran(Q)\,.
\tag\tgsimple
$$ 
This implies also that the RS coefficients are expressed explicitly by the 
formulas (\tggM) and (\tglaM). 
We rewrite the equalities (\tgsetIa) and (\tgsetIb) as 
$$
\gather 
\la=\bt\,\langle Vf,g\rangle\,, 
\tag\tgsetIIa \\
(\hK+\bt\,\hV-F-\la)g=-\bt\, Vf\,. 
\tag\tgsetIIb 
\endgather 
$$

Our task in this section is to solve the equation (\tgsetIIb), at least for 
particular values of $\bt$ and $\la$. The first observation is that 
(\tgsetIIb) can be reduced to the subspace $\Ran(P_S)$. We define 
$$
W\bl:=V(1+\bt\,\Gl P_RV)^{-1}=(1+\bt\,V\Gl P_R)^{-1}V \,.
\tag\tgW
$$ 
Using (\tgGla) and (\tgbound) we get an estimate valid for $|\la|<\w/2$, 
$$
\|\Gl P_R\|=\|(1-\la\,\GO P_R)^{-1}\GO P_R\|\le 
\left({\w\over2}-|\la|\right)^{-1}\,.
$$
Hence $W\bl$ is a well defined bounded operator and even analytically 
depending on $\bl$ in the domain 
$$
|\bt|\le{1\over12}\,\w\|V\|^{-1},\quad |\la|\le{1\over3}\,\w\,,
\tag\tgdomI
$$ 
and having the bound there 
$$
\|W\bl\|\le(1-|\bt| \|\Gl P_R\| \|V\|)^{-1} \|V\| \le 2\|V\|\,.
\tag\tgWV
$$
To simplify the notation we set 
$$
W_S\bl:= P_S\,W\bl\,P_S\,.
$$

\proclaim{Lemma 6.1}
If $g_S\in\Ran(P_S)\cap\Dom(\hK)$ solves the equation 
$$
\bigl(\hK+\bt\,W_S\bl -F-\la\bigr) g_S=-\bt\,P_S W\bl f\,,
\tag\tgred
$$
with $\bt$ and $\la$ being restricted by (\tgdomI), then 
$$
g=\bigl(1-\bt\,\Gl P_R W\bl\bigr) g_S- \bt\,\Gl P_R W\bl f
$$ 
belongs to $\Dom(\hK)$ and solves the equation (\tgsetIIb). 
\endproclaim 
\demo{Proof}
Obviously $g\in\Dom(\hK)$ since $\Ran(\Gl)=\Dom(\hK)$. Furthermore, 
$$
\align 
(\hK+\bt\,\hV -F-\la)\Gl P_R W\bl &= (P_R+\bt\,\hV\Gl P_R)W\bl \\
&= QV - P_S W\bl\,. 
\endalign 
$$ 
Hence 
$$
\align 
(\hK+\bt\,\hV -F-\la)g =&\, (\hK+\bt\,\hV -F-\la)g_S - \bt(QV- P_S W\bl)g_S \\
& \quad - \bt(QV- P_S W\bl)f \\
=&\, (\hK+\bt\,W_S\bl -F-\la)g_S+\bt\,P_S W\bl f-\bt\,Vf \\
=&\, -\bt\, Vf \,. \qed
\endalign 
$$
\enddemo 

We are about to solve the reduced equation (\tgred). Let us write, for 
the moment very formally, 
$$
(\hK+\bt\,W_S\bl -F-\la)^{-1}=(1+\bt\,\Gbl\,W_S\of\bl)^{-1}\Gbl
\tag 6.8 
$$
where 
$$
\Gbl:=(\hK+\bt\,W_S\di\bl -F-\la)^{-1}\,.
$$
Here we have used the obvious notation: $X\of:=X-X\di$ and $X\di$ is the 
diagonal part of an operator $X\in\BB(\Ran(Q))$. The next step is to justify 
the equality (6.8) in which the diagonal and off-diagonal parts of 
$W_S\bl$ have been separated. In order to treat the diagonal part we 
introduce another diophantine-like condition, this time in the parameters 
$\bt$ and $\la$, 
$$
|F_n-F-\la+\bt\,W\bl_{nn}|\ge\tpsi(n_2)\quad\fora\ n\in\SS,
\tag\tgdiobl
$$
with $\tpsi$ having been defined in (\tgpsi). If $\tau\ge\sg>1$ then, in 
virtue of (\tgdiow), the point $\bl=(0,0)$ obeys the condition (\tgdiobl). 
Let us rewrite (\tgdiobl) in an operator form. For this sake we define, 
parallelly to the definition of $D$ in (\tgD), a self-adjoint unbounded 
operator $L$ acting in $\KK\equiv l^2(\BZN)$ by 
$$
Lh_n=n_2\,h_n,\quad\forall h=(h_n)\in\Dom(L)\subset\KK\,.
\tag\tgL
$$
The condition (\tgdiobl) is equivalent to 
$$
\|\Gbl L^{-\tau}P_S\|\le {2\over\ga}\,.
\tag\tgdioop
$$

Let us now focus on the off-diagonal part of $W_S\bl$. First we prove an 
auxiliary estimate. 

\proclaim{Lemma 6.2} 
Suppose that $A$ is a bounded, densely defined operator in $\KK$, and 
$B,X\in\BB(\KK)$ are such that $[\,A,B\,]=0$, $\|B\| \|X\|<1$, the operators 
$\ad_A^{\ j}X$ are bounded for $1\le j\le p$, and 
$$
\|B\|\,\max_{1\le j\le p}\|\ad_A^{\ j}X\| \le 1\,.
$$
Then
$$
\|\ad_A^{\ p}X(1-BX)^{-1}\|\le 
\frac{p!\,(2^{p+1}-1)}{(1-\|B\| \|X\|)^{p+1}}\, 
\max_{0\le j\le p}\|\ad_A^{\ j}X\|\,.
$$
\endproclaim
\demo{Proof}
The case $p=0$ is evident. Suppose that $p\ge1$ and set temporarily 
$$
M:=\max_{0\le j\le p}\|\ad_A^{\ j}X\|\,.
$$
In virtue of Lemma 5.1 we have 
$$
\align
\|\ad_A^{\ p}X(1-BX)^{-1}\| &= 
\biggl\| \sum_{k=0}^\infty\ad_A^{\ p}X(BX)^k \biggr\| \\
&\le \sum\Sb \nu\in\BZ_+^p\\ \nu_1+2\nu_2+\dots+p\nu_p=p\endSb 
\ \frac{p!}{\prod\limits_{j=1}^p (j!)^{\nu_j}\,\nu_j!}\ 
\prod_{j=1}^p\|\ad_A^{\ j}X\|^{\nu_j} \\
&\phantom{\le}\qquad  \times\,\sum_{k=0}^\infty (k+1)k\dots(k+2-|\nu|)\, 
\|B\|^k\|X\|^{k+1-|\nu|} \\ 
&\le \sum\Sb \nu\in\BZ_+^p\\ \nu_1+2\nu_2+\dots+p\nu_p=p\endSb 
\ \frac{p!}{\prod\limits_{j=1}^p (j!)^{\nu_j}\,\nu_j!}\  M^{|\nu|}\  
\frac{\|B\|^{|\nu|-1} |\nu|!}{(1-\|B\| \|X\|)^{|\nu|+1}} \\ 
&\le \frac{p!\,M}{(1-\|B\| \|X\|)^{p+1}}\
\sum\Sb \nu\in\BZ_+^p\\ \nu_1+2\nu_2+\dots+p\nu_p=p\endSb 
\ \frac{|\nu|!}{\prod\limits_{j=1}^p (j!)^{\nu_j}\,\nu_j!} \,.
\endalign
$$
Here we have used that, for $|x|<1$ and $j\in\BZ_+$, 
$$ 
\sum_{k=0}^\infty k(k-1)\dots(k-j+1)\,x^{k-j} = 
\frac{j!}{(1-x)^{j+1}}\,.
$$
To finish the proof we estimate 
$$
\align
\sum\Sb \nu\in\BZ_+^p\\ \nu_1+2\nu_2+\dots+p\nu_p=p\endSb 
\ \frac{|\nu|!}{\prod\limits_{j=1}^p (j!)^{\nu_j}\,\nu_j!}
&< \sum_{k=0}^p \  \sum_{\nu\in\BZ_+^p,\,|\nu|=k} \binom k\nu
\, \prod_{j=1}^p \left({1\over j!}\right)^{\nu_j} \\
&= \sum_{k=0}^p \left(\sum_{j=1}^p {1\over j!}\right)^k \\
&< 2^{p+1}-1  \,.\quad\qed 
\endalign
$$
\enddemo 

Lemma 6.2 applied to $W\bl$ yields 
$$
\align
\|\ad_D^{\ r}W\bl\| &\le 
\frac{r!\,(2^{r+1}-1)}{(1-|\bt|\|\Gl P_R\|\|V\|)^{r+1}}\ 
\max_{0\le j\le r}\|\ad_D^{\ j}V\| \\
&\le r!\,2^{2r+2}\,\max_{0\le j\le r}\|\ad_D^{\ j}V\| 
\tag\tgadW
\endalign
$$
where the couple $\bl$ obeys (\tgdomI). 

In accordance with (\tgdecay), the existence of $\ad_D^{\ r}X$ implies a 
decay of the matrix entries of $X$. Below we derive some consequences of this 
fact. We consider also the situation when $X(z)$ is an analytic family of 
bounded operators. 

\proclaim{Lemma 6.3} 
Suppose that $A$ is a closed, densely defined operator in $\KK$, 
$\UU\subset\BC^N$ is open and $X(z)$, $z\in\UU$, is an analytic family of 
bounded operators such that $\Ran(X(z))\subset\Dom(A)$ for all $z\in\UU$.
If the family $AX(z)$ is locally uniformly bounded on $\UU$ then it is 
analytic. 
\endproclaim 
\demo{Proof} 
It is known (see VII\S1.1 in \cite{\Kato}) that a family of bounded operators 
$Y(z)$ is analytic if and only if it is locally uniformly bounded and there 
exist two
fundamental subsets $\XX_1,\XX_2\subset\KK$ such that the functions 
$\langle h_2,Y(z)h_1\rangle$ are analytic for all $h_1\in\XX_1$ and
$h_2\in\XX_2$. We apply this criterion to $Y(z)=AX(z)$, $\XX_1=\KK$
and $\XX_2=\Dom(A^\ast)$. Then the functions 
$$
\langle h_2,AX(z)h_1\rangle = \langle A^\ast h_2,X(z)h_1\rangle 
$$
are manifestly analytic. \qed 
\enddemo 
\flushpar
The symbol $\zeta(z)$ below stands for the Riemann zeta function, 
$$
\zeta(z):=\sum_{k=1}^\infty k^{-z}\,.
$$

\proclaim{Lemma 6.4} 
Suppose that $X\in\BB(\KK)$, $\ad_D^{\ r}X$ is bounded for some $r\in\BN$ 
and a number $\tau\in\BR$ satisfies $\tau\le r\al$. It holds true that 
\roster 
\item"(i)" $P_S Xf\in\Dom(L^\tau)$ and 
$$
\|L^\tau P_S Xf\|\le\sqrt{2\,\zeta(2r)}
\left({\w(1+\al)\over C_E}\right)^r\,
\max\{\|X\|,\,2^r\|\ad_D^{\ r}X\|\}\,,
\tag 6.13
$$ 
\item"(ii)" if $r\ge2$ then $\Ran(P_S X\of P_S)\subset\Dom(L^\tau)$ and 
$$
\|L^\tau P_S X\of P_S\|\le 2\,\zeta(r)
\left({\w(1+\al)\over C_E}\right)^r\,
\max\{\|X\|,\,2^r\|\ad_D^{\ r}X\|\}\,. 
\tag 6.14 
$$
\endroster 

Suppose, in addition, that $X(z)$ is an analytic family on an open set 
$\UU\subset\BC^N$ and $\ad_D^{\ r}X(z)$ is locally uniformly bounded. Then, 
otherwise under the same assumptions, the families $L^\tau P_S X(z)f$ and 
$L^\tau P_S X\of(z) P_S$ are analytic. 
\endproclaim 
\demo{Proof} 
The inequalities (6.13) and (6.14) follow readily from (\tgdecay) in 
combination with (\tgdistII) or (\tgdistI), respectively. For example, if 
$m,n\in\SS$, $m\not=n$, then 
$$
|(L^\tau X)_{mn}|\le\left({\w(1+\al)\over C_E}\right)^r\,
\max\{\|X\|,\,2^r\|\ad_D^{\ r}X\|\}\, m_2^{\tau-r\al}\,|m_2-n_2|^{-r}\,.
$$
Since $m_2^{\tau-r\al}\le 1$ and 
$$
\sum_{n\in\SS,\ n\not=m} |m_2-n_2|^{-r} \le 2\,\zeta(r)\,,
$$
the Schur-Holmgren criterion leads to (6.14). The verification of (6.13) is 
similar; instead of the Schur-Holmgren criterion  one uses the equality 
$$
\|L^\tau P_S Xf\|^2=\sum_{n\in\SS} |(L^\tau X)_{n\eta}|^2\,.
$$

Concerning the second part of the lemma, the inequalities (6.13) and (6.14) 
imply respectively that the families $L^\tau P_S X(z)f$ and 
$L^\tau P_S X\of(z) P_S$ are locally uniformly bounded on $\UU$ and so, in 
virtue of Lemma 6.3, they are analytic. \qed
\enddemo 

Now we can formulate an existence result. 

\proclaim{Proposition 6.5} 
Suppose that $\Vt\in C^r$, with $r\ge2$, and a couple $\bl\in\BR^2$ obeys the 
diophantine estimate (\tgdiobl), i.e., $\|\Gbl\,L^{-\tau}P_S\|\le 2/\ga$, 
with some $\tau$, $0\le\tau\le r\al$, and, in addition, it fulfills the 
inequalities 
$$
|\bt|\le\min\left\{C_g(r)^{-1},\,{1\over12}\,\w\|V\|^{-1}\right\},\quad 
|\la|\le{\w\over3}\,, 
\tag\tgdomr
$$
where
$$
C_g(r):={1\over\ga}\,32\,\zeta(r)\,r!\,
\left({8\w(1+\al)\over C_E}\right)^r\,
\max_{0\le j\le r} \|\ad_D^{\ j}V\| \,.
$$
Then the vector 
$$
g_S\bl:=-\bt\bigl(1+\bt\,\Gbl\,W_S\of\bl\bigr)^{-1}\Gbl\,P_S W\bl f
\tag\tgsols 
$$
is well defined and the vector 
$$
g\bl :=\bigl(1-\bt\,\Gl P_R W\bl\bigr) g_S\bl - \bt\,\Gl P_R W\bl f 
\tag\tgsol
$$ 
solves the equation (\tgsetIIb), i.e., 
$$
(\hK+\bt\,\hV-F-\la)g\bl=-\bt\, Vf \,.
\tag\tgsetIIbb
$$
\endproclaim
\demo{Proof} 
Recall the estimates (\tgWV) and (\tgadW), and note that (\tgdioop) implies 
$$
\Dom(L^\tau P_S)=\Ran(L^{-\tau}P_S)\subset\Dom(\Gbl)\,.
$$
According to Lemma 6.4 we have 
$$
P_S W\bl f\in\Dom(\Gbl),\quad \Ran(W_S\of\bl)\subset\Dom(\Gbl) \,,
$$ 
and it holds 
$$
\align 
\|\Gbl W_S\of\bl\| &\le \|\Gbl\,L^{-\tau}P_S\|\,\|L^\tau W_S\of\bl\| \\ 
&\le {2\over\ga}\cdot 2\zeta(r)
\left({\w(1+\al)\over C_E}\right)^r\,
\max\{\|W\bl\|,\,2^r\|\ad_D^{\ r}W\bl\|\} \\
&\le {1\over\ga}\,16\zeta(r)\,r!\, 
\left({8\w(1+\al)\over C_E}\right)^r\,
\max_{0\le j\le r}\|\ad_D^{\ j}V\| \\
&= {1\over2}\,C_g(r) \,.
\endalign
$$
This shows that $g_S\bl$ is well defined. 

Next we show that $g_S\bl$ solves (\tgred). It suffices to observe that \newline
$\Ran(\Gbl)\subset\Dom(\hK)$ and 
$$
\align
& (\hK+\bt\,W_S\bl-F-\la)\bigl(1+\bt\,\Gbl\,W_S\of\bl\bigr)^{-1}\Gbl P_S \\
&\quad= (\hK-F-\la+\bt\,W_S\di\bl) \\
&\quad\quad\times\
\biggl(1-\bt\,\Gbl\,W_S\of\bl
\bigl(1+\bt\,\Gbl\,W_S\of\bl\bigr)^{-1}\biggr)\Gbl P_S \\
&\quad\quad +\bt\,W_S\of\bl\bigl(1+\bt\,\Gbl\,W_S\of\bl\bigr)^{-1}\Gbl P_S \\
&\quad= P_S\,. 
\endalign
$$
Hence 
$$
(\hK+\bt\,W_S\bl-F-\la)g_S\bl=-\bt\,P_S\,W\bl f\,.
$$
The equality (\tgsetIIbb) is then a consequence of Lemma 6.1. \qed 
\enddemo

\bigpagebreak
\flushpar {\bf 7. More about the diophantine condition on $\bt$ and
$\la$}
\medpagebreak

The diophantine condition (\tgdiobl) involves the diagonal of the operator 
$W\bl$ whose definition (\tgW) represents in fact the geometric series  
$V-\bt\,V\Gl P_R V+\dots$. We start by checking more closely the term
$V\Gl P_R V$. Here is some additional notation. As one observes from 
(\tgV), a matrix entry $V_{mn}$ depends on $m_1$ and $n_1$ only through the 
difference $n_1-m_1$; we write 
$$
V_{mn}=:V(n_1-m_1,m_2,n_2).
$$
Clearly,
$$
\overline{V(k,p,q)}=V(-k,q,p).
$$
Set, for $n\in\SS$, 
$$
v_n(\la):=\sum_{k\in\BN}\,\frac{|V(k,n_2,n_2)|^2}{\w^2k^2-(F_n-F-\la)^2}\,. 
\tag\tgvn
$$ 
In virtue of the condition (\tgS), $v_n(\la)$ is well defined and even 
analytic for $|\la|\le\w/3$, with the uniform bound 
$$
|v_n(\la)| \le {\|V\|^2\over \w^2}\,\left(
\frac{1}{ 1-\left({5\over6}\right)^2}\,+ \sum_{k\ge2}{1\over k^2-1} 
\right) = 
{\|V\|^2\over \w^2}\,\left({36\over 11}+{3\over4}\right) \,.
$$
It is also clear that on this domain all derivatives of $v_n(\la)$ are 
bounded uniformly and independently of $n\in\SS$. 

\proclaim{Lemma 7.1} 
Suppose that $\Vt\in C^1$. Then there exists a constant $C_D>0$ such that the 
inequality 
$$
|(V\Gl P_R V)_{nn}+2(F_n-F-\la)\,v_n(\la)|\le C_D\,n_2^{\,-\al} 
\tag\tgbasic
$$
holds true for all $n\in\SS$ and all $\la\in\BR$, $|\la|\le\w/3$.
\endproclaim 
\demo{Proof} 
It suffices to verify (\tgbasic) for the indices $n\in\SS$ with sufficiently 
large components $n_2\in\BN$. So we assume that 
$$
1\le c\,n_2^{\,\al}\quad\text{where}\quad c:=C_E/3\w(1+\al)\,.
\tag 7.3 
$$
Write temporarily $\SS_\star:=\SS\cup\{\eta\}$. We express the diagonal 
element $(V\Gl P_R V)_{nn}$ as a sum, 
$$
(V\Gl P_R V)_{nn}=\sum_{m\not\in\SS_\star}|V_{nm}|^2\,(F_m-F-\la)^{-1}\,.
$$
Observe that the partial sum, with the summation index satisfying 
$m\not\in\SS$ and $m_2=n_2$, yields 
$$
\align
& \sum_{k\in\BZ,\ k\not=0} |V(k,n_2,n_2)|^2\,(\w k+F_n-F-\la)^{-1} \\
&\qquad = \sum_{k\in\BN} |V(k,n_2,n_2)|^2\,
\bigl( (\w k+F_n-F-\la)^{-1}+(-\w k+F_n-F-\la)^{-1}\bigr) \\ 
&\qquad = -2\,(F_n-F-\la)\,v_n(\la)\,.
\endalign
$$

We split the rest (with the summation index $m\not\in\SS_\star$, 
$m_2\not=n_2$) into two parts according to whether 
$|m_1-n_1|\ge c\,n_2^\al$ 
or $|m_1-n_1|< c\,n_2^\al$. In the first case we use the differentiability 
of $\Vt$, particularly the property 
$$
\sum_{m\in\BZN}|m_1-n_1|^2|V_{nm}|^2=\sum_{m\in\BZN}|(\ad_D V)_{nm}|^2 
\le \|\ad_D V\|^2\,,
$$
and the fact that 
$$
|F_m-F-\la|\ge |F_m-F|-|\la| \ge {\w\over2}-{\w\over3}={\w\over6} 
$$ 
holds true for $m\not\in\SS_\star$ and $|\la|\le\w/3$, to estimate 
$$
\align
\biggl|\sum \Sb m\not\in\SS_\star\\ |m_1-n_1|\ge c\,n_2^\al\endSb \,
|V_{nm}|^2\,(F_m-F-\la)^{-1}\biggr| &\le 
{6\over\w}\sum_{m\in\BZN}\left({|m_1-n_1|\over c\,n_2^\al}\right)^2\,
|V_{nm}|^2 \\
&\le {6\over\w c^2}\,\|\ad_D V\|^2\,n_2^{\,-2\al} \,.
\endalign
$$

In the second case we derive, using successively (\tgS), (\tgineq)
and (7.3), 
$$
\align 
|F_m-F-\la| &\ge |F_m-F_n|-|F_n-F|-|\la| \\
&\ge |E_{m_2}-E_{n_2}|-\w|m_1-n_1|-{\w\over 2}-{\w\over 3} \\
&> {C_E\over 1+\al}\,n_2^{\,\al} -\w c\,n_2^{\,\al} -\w \\ 
&\ge {C_E\over 3(1+\al)}\,n_2^{\,\al} \,.
\endalign 
$$ 
Hence 
$$
\align
\biggl|\sum 
\Sb m\not\in\SS_\star\\ |m_1-n_1|<c\,n_2^\al,\ m_2\not=n_2 \endSb \,
|V_{nm}|^2\,(F_m-F-\la)^{-1}\biggr| &\le 
{3(1+\al)\over C_E}\,n_2^{\,-\al} \sum_m |V_{nm}|^2 \\
&\le {3(1+\al)\over C_E}\,\|V\|^2\, n_2^{\,-\al} \,.
\endalign
$$
This completes the proof. \qed
\enddemo

Let us now define, for $n\in\SS$, 
$$
\align
& \tilde w_n\bl:=W\bl_{nn}-2\bt\,(F_n-F-\la)\,v_n(\la)\,, 
\tag\tgtwn \\
& w_n\bl:=\tilde w_n\bl/(1+2\bt^2\,v_n(\la))\,.
\tag\tgwn
\endalign 
$$
The diophantine estimate (\tgdiobl) can be rewritten as 
$$
|(F_n-F-\la)(1+2\bt^2\,v_n(\la))+\bt\,\tilde w_n\bl| \ge \tpsi(n_2)\quad 
\fora\ n\in\SS .
\tag 7.6  
$$
However, in the sequel we will replace (7.6) by a stronger condition, namely 
$$
|F_n-F-\la +\bt\,w_n\bl| \ge \tpsi(n_2)\quad 
\fora\ n\in\SS .
\tag\tgdiof
$$
Actually, (\tgdiof) implies (7.6) since from the expression (\tgvn) one 
finds readily that $v_n(\la)>0$ for all $n\in\SS$ and
all $\la\in\BR$, $|\la|\le\w/3$.

\proclaim{Lemma 7.2} 
The functions $w_n\bl$, $n\in\SS$, are analytic in a neighbourhood of the 
closed set 
$$
|\bt|\le{1\over12}\,\w\|V\|^{-1},\quad |\la|\le{1\over3}\,\w\,,
\tag\tgdomII 
$$ 
and on this set all their derivatives have bounds independent of
$n\in\SS$. 

Suppose, moreover, that $\Vt\in C^1$. Then for each $\veps>0$ there exist 
$k_\star\in\BN$ and $\delta_\star>0$ such that 
$$
\sup_{n_2\ge k_\star,\ |\bt|\le\delta_\star,\ |\la|\le\w/3}\,
|\pr_\bt w_n\bl| <\veps\,.
$$
\endproclaim 
\demo{Proof} 
Concerning the uniform boundedness, one deduces from the formulas (\tgtwn), 
(\tgwn) and from the properties of the functions $v_n(\la)$, as discussed 
above (see the definition (\tgvn)), that the problem reduces to an analogous 
assertion about the functions $W\bl_{nn}$, $n\in\SS$. But the latter case 
is quite obvious as the operator-valued function $W\bl$ is analytic in the 
indicated domain (see the definition (\tgW) and the related discussion). 

Again from the definition (\tgW) one finds that 
$$
W\bl=V-\bt\,V\Gl P_R V+\bt^2\,(V\Gl P_R)^2 W\bl.
$$
It follows readily that 
$$w_n\bl=V_{nn}-\bt\,\bigl((V\Gl P_R V)_{nn}+2(F_n-F-\la)\,v_n(\la)\bigr) 
+\bt^2\,\rho_n\bl \,,
\tag\tgexpand
$$
where $\rho_n\bl$, $n\in\SS$, are analytic functions on the same domain and 
with all derivatives bounded there independently of $n$. Lemma 7.1 then 
implies the result. \qed 
\enddemo

Denote by $\DD$ the closed set determined by the countable family of 
diophantine inequalities, 
$$
\DD:=\{\bl\in\BR^2;\ \bl\ \text{satisfies}\ (\tgdomII)\ \and\ 
(\tgdiobl)\}. 
\tag\tgDD
$$
In this definition the exponent $\tau$ (cf. (\tgpsi))
can be, in principle, any real number but
$\DD\not=\emptyset$ is possible only for $\tau\ge0$. Similarly, $\tD$ is 
defined in the same manner but with the condition (\tgdiobl) (or, 
equivalently (7.6)) being replaced by the stronger condition (\tgdiof). 
We know that if $\tau\ge\sg>1$ then $(0,0)\in\tD\subset\DD$. Next we are 
going to show that $\tD$ contains, and so does $\DD$, much more points than 
just the origin. But first we give two elementary lemmas.

\proclaim{Lemma 7.3} 
Suppose that $h\in C^2(\BR)$ and $h''(x)\ge a>0$ for all $x\in\BR$.
Then, for all $\veps>0$,
$$
|\{x\in\BR;\ |h(x)| < \veps\}|\le 4\,\sqrt{{\veps\over a}}\,.
$$
\endproclaim 
\demo{Proof} 
The function $h$ has exactly one local extreme, namely a minimal value 
$h_{\min}=h(x_{\min})$, and, according to whether $h_{\min}\ge\veps$ or 
$-\veps<h_{\min}<\veps$ or $h_{\min}\le -\veps$, the set 
$h^{-1}(]-\veps,\veps[)$ is either empty or an open bounded interval or a 
union of two open bounded intervals. Even in the case when 
$h^{-1}(]-\veps,\veps[)$ is an open interval we split it by the extremal 
point $x_{\min}$ into two intervals. So it suffices to estimate the measure 
of an interval $[\,x_1,x_2\,]$ such that 
$h([\,x_1,x_2\,])\subset[\,-\veps,\veps\,]$ and $h$ is monotone on 
$[\,x_1,x_2\,]$. For definiteness consider the case with $h$ increasing. 
Then $h'(x_1)\ge0$, $-\veps\le h(x_1)\le h(x_2)\le\veps$, and we have 
$$
\align 
2\veps &\ge \int_{x_1}^{x_2} h'(s)\, ds \\
&= (x_2-x_1)\,h'(x_1)+\int_{x_1}^{x_2} (x_2-s)\,h''(s)\, ds \\ 
&\ge \half\,(x_2-x_1)^2\,a \,.
\endalign 
$$
Hence $|x_2-x_1|\le 2\,\sqrt{\veps/a}$. \qed 
\enddemo 

\proclaim{Lemma 7.4} 
Suppose that $h\in C^2(\BR)$ and there are positive constants $a,b,c$ such 
that 
$$
|h(0)|\ge c,\quad |h'(0)|\le b,\quad \and\quad
|h''(x)|\ge a\quad  \fora\ x\in\BR.
$$
Then for all $\veps>0$, $\veps\le\min\{ b^2/a,c/2\}$, and all $\dl>0$ 
it holds true that 
$$
|\{x\in[\,-\dl,\dl\,];\ |h(x)|<\veps\}| \le 8\dl\,{b\over c}\, 
\sqrt{{\veps\over a}}\,.
$$
\endproclaim 
\demo{Proof} 
Let us assume for definiteness that $h''(x)\ge a$ for all $x\in\BR$. We
distinguish two cases. First, assume that $h(0)\ge c$ (and $c\ge 2\veps$). 
We apply Lemma 7.3 and the following observation. Consider the tangent line 
$y=h(0)+h'(0)\,x$ to the curve $y=h(x)$ and its intersection $(x_0,\veps)$ 
with the line $y=\veps$. If 
$h^{-1}([\,-\veps,\veps\,])\cap[\,-\dl,\dl\,]\not=\emptyset$ then, 
owing to the convexity, 
$$ 
\dl\ge|x_0|=|(h(0)-\veps)/h'(0)|\ge(c-\veps)/b\ge c/2b.
$$
This way we get 
$$
|h^{-1}(]-\veps,\veps[)\cap[\,-\dl,\dl\,]|\le 4\,\sqrt{{\veps\over a}}
\le  8\dl\,{b\over c}\,\sqrt{{\veps\over a}} \,.
$$

Second, assume that $h(0)\le-c$. Then the set $h^{-1}(]-\veps,\veps[)$ 
is a union of two open bounded intervals. Consider, for example, that one on 
which $h$ is increasing and denote it by $]x_1,x_2[$. If 
$[\,x_1,x_2\,]\cap[\,-\dl,\dl\,]\not=\emptyset$ then $0<x_1\le\dl$ and, 
of course, $h(x_1)=-\veps$, $h(x_2)=\veps$. By convexity we have 
$$
{2\veps\over x_2-x_1}={h(x_2)-h(x_1)\over x_2-x_1}\ge 
{h(x_1)-h(0)\over x_1} \ge {c-\veps\over\dl} \ge {c\over 2\dl} 
$$
and so $|x_2-x_1|\le4\dl\veps/c$. But the restriction $\veps\le b^2/a$ 
implies 
$$
|h^{-1}(]-\veps,\veps[)\cap[\,-\dl,\dl\,]|\le 8\dl\,{\veps\over c}  
\le 8\dl\,{b\over c}\, \sqrt{{\veps\over a}}\,. \quad \qed
$$
\enddemo 

The following proposition gives a characterization of the set $\DD$ 
which is determined, according to (\tgDD),
by the diophantine-like condition (\tgdiobl). 

\proclaim{Proposition 7.5} 
Suppose that $\Vt\in C^1$ and the exponents $\tau$ and $\sg$ in (\tgpsi) 
satisfy $\sg>1$ and $\tau>2\sg+2$. Furthermore, suppose that 
$\vp\in C^2(\BR)$, $\vp(0)=\vp'(0)=0$ and $\vp''(0)\not=0$. Set 
$$
I(\vp):=\left\{\bt\in\BR;\ |\bt|\le{1\over12}\,\w\|V\|^{-1},\ 
|\vp(\bt)|\le{\w\over3}\quad\and\quad (\bt,\vp(\bt))\in\DD\right\}\,.
\tag\tgIphi
$$
Then 0 is a point of density of the set $I(\vp)$, i.e., 
$$
\lim_{\dl\downarrow 0}\,{1\over 2\dl}\,|I(\vp)\cap [\,-\dl,\dl\,]|=1 \,.
$$
\endproclaim 
\demo{Proof} 
Set (in this proof)
$$
h_n(\bt):=F_n-F-\vp(\bt)+\bt\,w_n(\bt,\vp(\bt)),\quad n\in\SS.
$$
For $\dl>0$ sufficiently small we have, as $\tD\subset\DD$, 
$$
\align 
& [\,-\dl,\dl\,]\setminus I(\vp)\subset
\bigcup_{n\in\SS}\Phi_n(\dl)\,,\quad\text{where}\\ 
& \Phi_n(\dl):=\{\bt\in[\,-\dl,\dl\,];\ |h_n(\bt)|<\tpsi(n_2)\}\,.
\endalign 
$$
One finds that (cf. (\tgexpand)) 
$$
\align 
& |h_n(0)|=|F_n-F|\ge \psi(n_2),\ h_n'(0)=w_n(0,0)=V_{nn}, \\
& h_n''(\bt)=-\vp''(\bt)+2\,\pr_\bt w_n(\bt,\vp(\bt))+ 
2\,\pr_\la w_n(\bt,\vp(\bt))\,\vp'(\bt)+O(|\bt|)\,.
\endalign 
$$
From Lemma 7.2 and from the fact that $\vp'(0)=0$ we conclude that there 
exist $k_\star\in\BN$ and $\dl_\star>0$ such that 
$$
|h_n''(\bt)|\ge a,\quad\forall n\in\SS,\ n_2\ge k_\star,\ \and\
\forall\bt\in[\,-\dl_\star,\dl_\star\,],
$$
where 
$$
a:=|\vp''(0)|/2\,. 
\tag 7.12 
$$
Naturally we choose $\dl_\star>0$ sufficiently small so that the
inequalities $|\bt|\le\w/(12\|V\|)$ and $|\vp(\bt)|\le\w/3$ are
fulfilled for $|\bt|\le\dl_\star$.

Furthermore, since $\psi(k)>\tpsi(k)$, $\forall k\in\BN$, there exists
a sequence of positive numbers, $\{\bt_n\}_{n\in\SS}$, such that 
$0<\bt_n\le \dl_\star$ and 
$$
|h_n(\bt)|\ge\tpsi(n_2),\quad\forall\bt\in[\,-\bt_n,\bt_n\,],\
\forall n\in\SS\,.
$$
In other words, $\Phi_n(\dl)=\emptyset$ for $\dl\le\bt_n$. If
necessary we increase the value $k_\star\in\BN$ so that
$$
\tpsi(k)\le \|V\|^2/a,\quad \forall k\ge k_\star\,.
\tag 7.13 
$$
Now we can apply Lemma 7.4, with $c=\psi(n_2)$, $b=\|V\|$, 
$a$ given in (7.12) and 
$\veps=\tpsi(n_2)$, to the set $\Phi_n(\dl)$. If $n_2\ge k_\star$ then the 
assumption $\veps\le\min\{b^2/a,\,c/2\}$ is satisfied owing to (7.13) and 
to the fact that $\psi(k)\ge 2\,\tpsi(k)$, $\forall k\in\BN$. Hence
$$
|\Phi_n(\dl)|\le 8\dl\,{\|V\|\over\sqrt{a}}\,
{\sqrt{\tpsi(n_2)}\over\psi(n_2)}\le\const\ 2\dl\,n_2^{\,-\half\tau+\sg}\,.
$$

Summing up, provided 
$$
0<\dl\le\min_{n\in\SS,\ n_2<k_\star}\bt_n\quad\and\quad\dl\le\dl_\star
$$
(which implies that $\Phi_n(\dl)=\emptyset$ for $n_2<k_\star$) we have
the estimate
$$
\align 
{1\over 2\dl}\,|[\,-\dl,\dl\,]\setminus I(\vp)| &\le 
{1\over 2\dl}\,\biggl|\bigcup_{n\in\SS,\ n_2\ge k_\star} 
\Phi_n(\dl)\biggr| \\
&\le \const\ \sum_{n\in\SS,\ \bt_n < \dl} n_2^{\,-\half\tau+\sg}\,. 
\endalign 
$$
Recall that the projection $\SS\to\BN\setminus\{\eta_2\}$ is one-to-one. 
Hence the sum
$\sum_{n\in\SS} n_2^{\,-\half\tau+\sg}$ converges. Since 
$$\bigcap_{\dl>0}\{ n\in\SS;\ \bt_n<\dl\}=\emptyset
$$
we get 
$$
\lim_{\dl\downarrow 0}\,{1\over 2\dl}\,
|[\,-\dl,\dl\,]\setminus I(\vp)|=0\,. \quad\qed
$$
\enddemo 

\bigpagebreak
\flushpar {\bf 8. Implicit equation, completion of the proof}
\medpagebreak

Let us return to Proposition 6.5. Suppose that $\Vt\in C^r$, with $r\ge2$, 
and that $\tau\le r\al$, and denote by $\DD(r)$ the intersection of the 
set $\DD$ defined in (\tgDD) with the closed unit ball in $\BR^2$ and with 
the closed set determined by the inequalities (\tgdomr). In fact,
$\DD(r)$, as well as $\DD$, depends also on the exponent $\tau$,
$\tau\ge 0$, (cf. (\tgpsi)).
Then for all $\bl\in\DD(r)$ the vector $g\bl$
defined in (\tgsols) and (\tgsol) solves the equation (\tgsetIIb). 
Recall that $\hV=QVQ$; consequently 
$$
Vh=\hV h+\langle Vf,h\rangle\,f,\quad\forall h\in\Ran(Q)\,.
$$
Altogether this means that 
$$
(K+\bt V)g\bl=(F+\la)g\bl+\bt\langle Vf,g\bl\rangle\,  f-\bt Vf\,.
$$
Since $Kf=Ff$ we arrive at the equality 
$$ 
(K+\bt V)(f+g\bl)=(F+\la)(f+g\bl)+(G\bl -\la)f
\tag\tgalmost
$$
where 
$$
G\bl:=\bt\,\langle Vf,g\bl\rangle\,. 
\tag\tgG 
$$ 
Thus our final task, in order to get an eigen-value and an eigen-vector, 
is to solve the implicit equation 
$$
\la-G\bl=0\,,
\tag\tgimpl
$$
which is nothing but the equation (\tgsetIIa). 

We will solve (\tgimpl) in a Lipschitz class. The notion of Lipschitz 
functions as well as their properties needed for our purposes are recalled 
in Appendix C. This also concerns the celebrated Whitney extension theorem 
\cite{\Stein}. We remind the reader that the target space is generally 
allowed to be a Banach space or, more particularly, a Banach algebra.
When indicating that a function belongs to a Lipschitz class supported on
a closed set we
always assume tacitly that this concerns the corresponding
restriction. We have to decide about the Lipschitz property of the 
vector-valued function $g\bl$ defined on $\DD(r)$. Looking at the formulas 
(\tgsols) and (\tgsol) one finds immediately that $\Gbl$ is the only 
operator-valued function occurring in the expressions which is not analytic 
(and so automatically Lipschitz). 

\proclaim{Lemma 8.1} 
For all $\ell\in\BZ_+$, the function $\Gbl\,L^{-\tau(\ell+2)} P_S$ belongs 
to the Lipschitz class $\Lip(\ell+1,\DD\cap\bar B_1)$ where 
$\bar B_1\subset\BR^2$ is the closed unit ball. 
\endproclaim 
\demo{Proof} 
Set temporarily 
$$
\phi_n\bl:= F_n-F-\la+\bt\,W\bl_{nn},\quad n\in\SS;
$$ 
hence $\Gbl_{mn}=\phi_m\bl^{-1}\,\dl_{mn}$. Owing to (\tgbound), the 
operator-valued function $(\hK-F-\la+\bt\,W\di\bl)P_S$ is bounded and 
analytic on a neighbourhood of the closed set determined by (\tgdomII), 
and so it belongs to $\Lip(\ell+1,\DD\cap\bar B_1)$; denote by $M_\ell$ 
its Lipschitz norm. This implies that ($M(\cdot)$ stands for the Lipschitz 
norm) 
$$
\phi_n\in\Lip(\ell+1,\DD\cap\bar B_1)\quad\and\quad M(\phi_n)\le M_\ell 
\quad\fora\ n\in\SS\,.
$$ 
Since $|\phi_n\bl|\ge(\ga/2)\,n_2^{\,-\tau}$ (cf. (\tgdiobl))  
one can apply Proposition C.5, with the constant $C_L(2,\ell)$ redenoted as 
$C(\ell)$, to conclude that 
$$
M(\phi_n\bl^{-1})\le C(\ell)\,M_\ell^{\,\ell+1}
\left({2\over\ga}\,n_2^{\,\tau}\right)^{\ell+2}\,,\ \forall n\in\SS.
$$
This completes the proof for 
$$
M(\Gbl\,L^{-\tau(\ell+2)} P_S)\le C(\ell)\,M_\ell^{\,\ell+1}
\left({2\over\ga}\right)^{\ell+2} < \infty\,. \quad\qed
$$
\enddemo 

\proclaim{Lemma 8.2} 
Suppose that $\Vt\in C^r$, with $r\ge2$ and 
$0\le\tau(\ell+2)\le r\al$, and 
$\ell\in\BZ_+$. Then the vector-valued function $g\bl$ defined in (\tgsol) 
belongs to the class $\Lip(\ell+1,\DD(r))$. 
\endproclaim 
\demo{Proof} 
The function $\Gl P_R W\bl$ is analytic in a neighbourhood of $\DD(r)$ and 
so it belongs to the Lipschitz class of any order. Hence, in virtue of the 
relation (\tgsol) and Proposition C.4, it suffices to verify the assertion 
for the function $g_S\bl$ instead of $g\bl$. Here the Banach algebra in 
question is $\BB(\KK)$. The fact that the expressions involve also
$\KK$-valued functions does not mean a serious complication:
either one can modify, in an obvious way,
Proposition C.4 or one can replace everywhere vectors $h\in\KK$ by the 
rank-one operators $\tilde h\in\BB(\KK)$, 
$\tilde hx:=\langle f,x\rangle h$ 
(e.g., $f$ would be replaced by $P$). Furthermore, from Lemma 6.4 we 
deduce that the functions $L^{\tau(\ell+2)}\,W_S\of\bl$ and 
$L^{\tau(\ell+2)}\,P_S W\bl f$ are analytic as well. Checking the formula 
(\tgsols) one concludes readily from Lemma 8.1, Proposition C.4 
and Proposition C.5 that $g_S\bl$ belongs indeed to the indicated 
Lipschitz class. \qed 
\enddemo

Let us add a remark to Lemma 8.2. From the proof and from the formulas 
(\tgsols), (\tgsol) it is quite obvious that the functions 
$\bt^{-1}\,g_S\bl$ and $\bt^{-1}\,g\bl$ belong to $\Lip(\ell+1,\DD(r))$. 
If $\tau\ge\sg$ then $(0,0)\in\DD(r)$ and we have 
$$
\align 
& \bt^{-1}\,g_S\bl\bigl|_{\bl=(0,0)} = -\GO P_S Vf\,, \\
& \bt^{-1}\,g\bl\bigl|_{\bl=(0,0)} = -\GO P_S Vf-\GO P_R Vf=-\GO Vf \,.
\endalign
$$
The set $\DD(r)$ is closed and so we can apply the Whitney extension 
theorem to the function $\bt^{-1}\,g\bl$. As a consequence we get an 
extension $\tilde g\bl\in\Lip(\ell+1,\BR^2)$ of the function $g\bl$ 
itself. Then, according to the formula (\tgG), the function 
$G\bl\in\Lip(\ell+1,\DD(r))$ as well and 
$$
\tilde G\bl:=\bt\,\langle Vf,\tilde g\bl\rangle\in\Lip(\ell+1,\BR^2) 
\subset C^\ell(\BR^2) 
$$
is an extension of it. Moreover, the previous remark implies that the 
function $\bt^{-2}\,\tilde G\bl$ belongs to the class 
$\Lip(\ell+1,\BR^2)$, too. Consequently, (if $\tau\ge\sg$) 
$$
\pr_\la^j \tilde G(0,0)=0\quad\and\quad 
\pr_\bt\pr_\la^k \tilde G(0,0)=0\quad\for\ j,k\in\BZ_+,\ j\le\ell\ \and\ 
k\le\ell-1, 
\tag\tgderG
$$
and, if $\ell\ge2$, 
$$
\pr_\bt^2 \tilde G(0,0) = 2\bt^{-2}\,\tilde G\bl\biggl|_{\bl=(0,0)} = 
-2\,\langle Vf,\GO Vf\rangle \,.
$$

Suppose that $\ell\ge1$. Instead of (\tgimpl) we shall consider the implicit 
equation in $\BR^2$, this is to say with the extended function 
$\tilde G\in C^\ell(\BR^2)$, 
$$
\la-\tilde G\bl=0. 
\tag\tgimplex 
$$
Since 
$$
\la-\tilde G\bl\biggl|_{\bl=(0,0)}=0,\quad 
\pr_\la(\la-\tilde G\bl)\biggl|_{\bl=(0,0)}=1, 
$$
the implicit mapping theorem guarantees the existence of $\bt_\star>0$ and 
of a unique function $\tilde\la\in C^\ell([\,-\bt_\star,\bt_\star\,])$ such 
that 
$$
\tilde\la(0)=0\quad\and\quad
\tilde\la(\bt)=\tilde G(\bt,\tilde\la(\bt))\quad\fora\ \bt\in
[\,-\bt_\star,\bt_\star\,]\,.
\tag\tgsolvex 
$$
Let us calculate the lowest order derivatives of $\tilde\la$: 
$$
\tilde\la'(\bt)=\bigl(1-\pr_\la\tilde G(\bt,\tilde\la(\bt))\bigr)^{-1} 
\pr_\bt\tilde G(\bt,\tilde\la(\bt))\,,
\tag\tgderla 
$$
and so $\tilde\la'(0)=0$. If $\ell\ge2$ then 
$$
\tilde\la''(0)=\pr_\bt^2\tilde G(0,0)=-2\,\langle Vf,\GO Vf\rangle\,. 
\tag\tgderlaII
$$

\proclaim{Proposition 8.3} 
Suppose that $\Vt\in C^r$, with $r\ge2$ and $\tau(\ell+2)\le r\al$, and 
$\tau\ge\sg>1$, $\ell\in\BN$. Then there exist $\bt_\star>0$ 
and a solution 
$\tilde\la\in\Lip(\ell+1,[\,-\bt_\star,\bt_\star\,])$ of the implicit 
equation (\tgimplex), i.e., the equalities (\tgsolvex) hold. Furthermore, 
the $\Ran(Q)$-valued function $\tilde g(\bt,\tilde\la(\bt))$, too, belongs 
to the class $\Lip(\ell+1,[\,-\bt_\star,\bt_\star\,])$. 

If, for some $\bt\in[\,-\bt_\star,\bt_\star\,]$, 
$(\bt,\tilde\la(\bt))\in\DD$ then $F+\tilde\la(\bt)$ is an eigen-value of 
$K+\bt V$ corresponding to the eigen-vector 
$f+g(\bt,\tilde\la(\bt))$.
\endproclaim 
\demo{Proof} 
We already know that $\tilde\la\in C^\ell([\,-\bt_\star,\bt_\star\,])$. 
To complete the proof we have to show that $\tilde\la$ even belongs to
$\Lip(\ell+1,[\,-\bt_\star,\bt_\star\,])$ or, equivalently, \newline
$\tilde\la^{(\ell)}\in\Lip(1,[\,-\bt_\star,\bt_\star\,])$. Let us first 
specify more precisely the choice of $\bt_\star>0$. We can assume, 
because of (\tgderG), that 
$$
|\pr_\la\tilde G(\bt,\tilde\la(\bt))|\le\half,\quad
\forall\bt\in[\,-\bt_\star,\bt_\star\,]\,.
$$ 
Furthermore, since $\tilde\la(0)=0$, we require the points 
$(\bt,\tilde\la(\bt))$, with $\bt\in[\,-\bt_\star,\bt_\star\,]$, to satisfy 
the inequalities (\tgdomr) and, at the same time, to belong to the unit
ball $\bar B_1$. In other words, if $\bt\in[\,-\bt_\star,\bt_\star\,]$ 
and $(\bt,\tilde\la(\bt))\in\DD$ then $(\bt,\tilde\la(\bt))\in\DD(r)$. 

In virtue of (\tgderla) we have 
$$
\sum_{j=1}^\ell \binom{\ell-1}{j-1}\,\tilde\la^{(j)}(\bt)\, 
{d^{\ell-j}\over d\bt^{\ell-j}}
\bigl(1-\pr_\la\tilde G(\bt,\tilde\la(\bt))\bigr)=
{d^{\ell-1}\over d\bt^{\ell-1}}\,\pr_\bt\tilde G(\bt,\tilde\la(\bt))\,.
\tag 8.9 
$$
Deduce from Proposition C.6 and from the fact that 
$\tilde\la\in\Lip(\ell,[\,-\bt_\star,\bt_\star\,])$ that 
$$
\pr_\bt^j\pr_\la^k\tilde G(\bt,\tilde\la(\bt))\in 
\Lip(\ell-j-k+1,[\,-\bt_\star,\bt_\star\,])\quad\text{if}\ 
1\le j+k\le\ell.
$$
One can express $\tilde\la^{(\ell)}$ from the identity (8.9); according 
to our choice of $\bt_\star$, 
\newline
$|1-\pr_\la\tilde G(\bt,\tilde\la(\bt))|\ge 1/2$. Now the usual rules 
of differentiation jointly with Proposition C.5 and Proposition C.4 imply 
that $\tilde\la^{(\ell)}\in\Lip(1,[\,-\bt_\star,\bt_\star\,])$. 

This is also because of Proposition C.6 that we can claim that the composed 
function $\tilde g(\bt,\tilde\la(\bt))$ belongs to 
$\Lip(\ell+1,[\,-\bt_\star,\bt_\star\,])$. The final part of the assertion  
can be seen immediately from the equality (\tgalmost) for it holds, by our 
choice of $\bt_\star$ specified above: if 
$\bt\in[\,-\bt_\star,\bt_\star\,]$ and $(\bt,\tilde\la(\bt))\in\DD$ then 
$$
\tilde\la(\bt)=\tilde G(\bt,\tilde\la(\bt))=G(\bt,\tilde\la(\bt))\,.
\quad\qed 
$$
\enddemo

\demo{Proof of Theorem 2.1} 
The first part of the theorem has been already proven in Section 5 -- 
see Remark (2) at the end of the section. All the steps needed to show the 
second part, too, have been already stated and so we have just to summarize 
them. We make the choice of $\sg$ and $\tau$ as specified in (\tgexpo). 
Proposition 6.5 guarantees the existence of a solution $g\bl$ of the 
eigen-vector equation (\tgsetIIb) provided $\bl$ belongs to $\DD(r)$, a 
closed set introduced in the beginning of this section. Consider now the 
function $\tilde\la\in\Lip(\ell+1,[\,-\bt_\star,\bt_\star\,])$, as
described in Proposition 8.3. Set 
$$
I:=[\,-\bt_\star,\bt_\star\,]\cap I(\tilde\la)\,,
$$
with $I(\vp)$ having been defined in (\tgIphi). Denote by $\Fb$ the 
restriction of the function $F+\tilde\la(\bt)$ to the set $I$ and by 
$\fb$ the restriction of $f+\tilde g(\bt,\tilde\la(\bt))$ to the same set. 
According to Proposition 8.3, 
$$
(K+\bt V)\fb=\Fb\fb\quad\fora\ \bt\in I\,.
$$
Since $\ell$, as specified in Theorem 2.1, fulfills $\ell\ge2$, and since 
$\tilde\la(0)=\tilde\la'(0)=0$, $\tilde\la''(0)= 2\la_2\neq 0$ 
(cf. (\tgderla) and (\tgderlaII)), Proposition 7.5 tells us that 0 is a 
point of density of $I$. Finally, we know, again from Proposition 8.3, that 
both $\Fb$ and $\fb$ belong to the Lipschitz class $\Lip(\ell+1,I)$. 
According to Lemma 4.1, the same is true for $(K+\bt V)\fb$. Moreover,
since
$g\bl\in\Ran(Q)$ we have $\langle f,\fb\rangle=1$ for all $\bt\in I$.
Then, as explained in Section 4, the coefficients from the asymptotic 
expansion of the functions 
$\Fb$ and $\fb$ at $\bt=0$ obey the equations (\tgeqI) 
(or, equivalently (\tgeqII)). To complete the proof we note that 
Proposition 5.4 ensures the existence and uniqueness 
of the solution to this system of equations and Proposition 4.3 gives 
its explicit form coinciding with the standard 
formulas known for RS series. 
\enddemo 

\bigpagebreak
\flushpar {\bf Appendix A. Density of the spectrum for almost all
frequencies}
\medpagebreak

\proclaim{Proposition A.1} 
Suppose that a set $\EE\subset\BR$ fulfills $\sup\EE=+\infty$. Then the set 
$\w\BZ+\EE$ is dense in $\BR$ for almost all $\w\in\BR$ 
(in the Lebesgue sense). 
\endproclaim 

As $-\BZ=\BZ$ we can consider only positive values of $\w$. Furthermore,
we make use of the facts that the positive half-line can be covered by a 
countable union of open bounded intervals and that the countable system of 
open intervals with rational endpoints forms a basis of the topology in 
$\BR$. We conclude from this that the following proposition, seemingly 
weaker, is in fact equivalent to Proposition A.1. 

\proclaim{Proposition A.2} 
Suppose that we are given an open interval $]a,b[$, $0<a<b<\infty$, and a 
compact interval $[\,u,v\,]$. Then, under the same assumptions about the
set $\EE$ as in Proposition A.1, it holds
$$
(\w\BZ+\EE)\cap[\,u,v\,]\neq\emptyset\quad
\text{for almost all } \w\in\, ]a,b[\,.
$$
\endproclaim 

\proclaim{Lemma A.3} 
Suppose that $\EE$ is the same as in Proposition A.1, $[\,u,v\,]$ is a 
compact interval, $\UU\subset\,]v-u,+\infty[$ is an open set and 
$|\UU|<\infty$. Then there exists $x_\star\in\BR$ such that for all 
$x>x_\star$ one can find a closed set $\MM(x)\subset\UU$ with the 
properties: 
\roster 
\item 
$\dsize{
(x+\w\BZ)\cap[\,u,v\,]\neq\emptyset\quad\fora\ \w\in\MM(x)\,,
}$ 
\item 
$\dsize{
|\MM(x)|\ge{1\over4}\,(v-u)\dsize\int_\UU{1\over s}\,ds\,. 
}$
\endroster 
\endproclaim 
\demo{Proof} 
$\UU$, as an open set, is at most countable disjoint union of open 
intervals. Since 
$$
\int_\UU{1\over s}\,ds\le{1\over v-u}\,|\UU|<\infty 
$$
there exists a finite subunion $\UU'=\bigcup\UU_i\subset\UU$, formed 
necessarily by bounded intervals, such that 
$$
\int_{\UU'}{1\over s}\,ds=\sum_i \int_{\UU_i}{1\over s}\,ds \ge
\half \int_\UU{1\over s}\,ds\,.
$$
We will seek a family of closed subsets $\MM_i(x)\subset\UU_i$ so that, 
for each $i$, the properties (1) and (2) are valid for $\MM_i(x)$ and 
$\UU_i$ in the place of $\MM(x)$ and $\UU$, respectively, with the only 
difference: we replace the factor 1/4 in (2) by 1/2. Suppose that we are 
successful. Then the disjoint union $\MM(x):=\bigcup_i \MM_i(x)$ has all 
the required properties. 

Fix an index $i$ and write $\UU_i=\,]a,b[$ where $0<v-u\le a<b<\infty$. 
Assume that 
$$
x>\max\left\{0,v,{vb-ua\over b-a}\right\}\,.
$$
Then
$$
0<{x-u\over b}<{x-v\over a}\le{x-v\over v-u} 
$$
and the union 
$$ 
\MM_i(x):=\bigcup\Sb k\in\BN \\ (x-u)/b <k< (x-v)/a \endSb 
\left[{x-v\over k},{x-u\over k}\right] 
$$
is disjoint. Consequently, 
$$
\align 
|\MM_i(x)| &= \sum_{ (x-u)/b <k< (x-v)/a  } {v-u\over k} \\
&= (v-u)\,\log{b\over a} +O(x^{-1}) \\
&\ge \half\,(v-u)\,\log{b\over a} 
\endalign 
$$
for sufficiently large $x$. Moreover, if $\w\in\MM_i(x)$ then there exists 
$k\in\BN$ such that $x-\w k\in[\,u,v\,]$. \qed 
\enddemo 

\demo{Proof of Proposition A.2} 
Clearly, $(x+\w\BZ)\cap[\,u,v\,]\neq\emptyset$ for all $\w$,
$0<\w\le v-u$, and all $x\in\BR$. Consequently we can assume, without
loss of generality, that $v-u\le a$. Using Lemma A.3 we construct 
successively a sequence $\MM(x_1),\MM(x_2),\dots$ formed by disjoint closed 
subsets of the interval $]a,b[$, with the points $x_k\in\EE$, so that 
$\MM(x_k)$ is related to the open set
$\UU_k=\, ]a,b[\,\setminus\NN_k$ where
$$
\NN_1:=\emptyset\quad\and\quad\NN_k:=\MM(x_1)\cup\dots\cup\MM(x_{k-1}) 
\quad\for\ k\ge2\,.
$$
Set 
$$
\NN:=\bigcup_{k\in\BN}\NN_k=\bigcup_{k\in\BN}\MM(x_k)\,. 
$$ 
The property (1) implies 
$$
(\w\BZ+\EE)\cap[\,u,v\,]\neq\emptyset,\quad\forall\w\in\NN\,.
$$
Furthermore, $|\NN|=\lim |\NN_k|$ and, owing to the property (2), we have 
$$
|\NN_{k+1}|-|\NN_k|=|\MM_k|\ge{1\over4}\int_{]a,b[\,\setminus\NN_k} 
\,{1\over s}\, ds\,. 
\tag A.1 
$$
Passing in (A.1) to the limit $k\to\infty$ we get 
$$
0\ge \int_{]a,b[\,\setminus\NN} \,{1\over s}\, ds 
$$
and so $|]a,b[\,\setminus\NN|=0$. \qed 
\enddemo 

\bigpagebreak
\flushpar {\bf Appendix B. A perturbation without
Rayleigh-Schr\"odinger series}
\medpagebreak

In this appendix we exhibit an example of a perturbation for which
$\lambda_2$ given in (\tglam) does not exist.
The symbols $H,K,E_k,F_n,e_k,f_n$ and $V$ retain their meaning from 
Section 2. However we don't require anymore that the eigen-values $E_k$ of 
the Hamiltonian $H$ obey the gap condition (\tggap). Instead we impose 
another restriction which has this time a multiplicative form. More 
precisely, we assume that there exist constants $C_M>0$ and $\mu>0$ 
such that 
$$
j<k\Longrightarrow {E_k\over E_j}\ge C_M\left({k\over j}\right)^\mu\,.
\tag B.1 
$$ 
Generally speaking, the conditions (\tggap) and (B.1) are independent. 
However in some cases, for example when the eigen-values $E_k$ grow 
polynomially, $E_k=\const\,k^\mu$, the condition (B.1) appears to be milder 
than (\tggap). Actually, the condition (\tggap) is satisfied provided
$\mu>1$ while (B.1) holds obviously for any $\mu$ positive. 

\proclaim{Proposition B.1} 
Suppose that the spectrum $\Spec(H)$ satisfies the condition (B.1).
Then, for almost all $\w>0$,
there exists a bounded self-adjoint perturbation $\Vt$
which is a $2\pi$-periodic and strongly continuous function of $t$ and 
such that the Rayleigh-Schr\"odinger coefficient $\la_2$ given in (\tglam) 
doesn't exist, i.e., the series 
$$
\sum_{n\in\BZN,\ n\neq\eta} {1\over F_n-F_\eta}\,|V_{n\eta}|^2
\tag B.2
$$
diverges, and this holds true for all $\eta\in\BZN$.
\endproclaim 

Let us introduce yet another condition. Namely, one requires that there
exist constants $C_M>0$ and $\mu>0$ such that
$$
j<k\Longrightarrow {E_k\over E_j}\ge C_M\,
\exp\bigl(\mu(k-j)\bigr)\,.
\tag B.3
$$ 
Since, for $1\le j\le k$, it is true that $k-j\ge\log k-\log j$, (B.3)
implies (B.1). However we shall show that, in the text of
Proposition B.1, one can replace harmlessly, without doing any other
change, "the condition (B.1)" by "the condition (B.3)". The new
proposition will be called {\it Proposition B.1 (modified)}.

\proclaim{Lemma B.2}
Suppose that the spectrum of a Hamiltonian $H$ satisfies the condition
(B.1). Then $H$ can be decomposed into a direct sum,
$H=\sum^\oplus_{a\in\BZ_+} H_a$, so that the spectrum of each summand
satisfies the condition (B.3), of course, with modified constants
$C_M>0$ and $\mu>0$.
\endproclaim

\demo{Proof}
Each $j\in\BN$ can be written in a unique way as $j=a+2^k$, with
$a,k\in\BZ_+$ and $a\le 2^k-1$. For a given $a\in\BZ_+$, denote by
$\kappa(a)$ the smallest non-negative integer such that
$a\le 2^{\kappa(a)}-1$, and set
$$
N(a):=\{a+2^{\kappa(a)+k-1};\ k\in\BN\}\,.
$$
According to what we have said,
$$
\BN=\bigcup_{a\in\BZ_+}N(a)
$$
is a disjoint union. It induces a decomposition of $H$,
$H=\sum^\oplus_{a\in\BZ_+} H_a$, so that
$\Spec(H_a)=\{E_k;\ k\in N(a)\}$. Furthermore, if $j,k\in\BN$ and $j<k$
then
$$
C_M \left(\frac{a+2^{\kappa(a)+k-1}}{a+2^{\kappa(a)+j-1}}\right)^\mu
\ge C_M
\left(\frac{2^{\kappa(a)+k-1}}{(a+1)2^{\kappa(a)+j-1}}\right)^\mu
={C_M\over (a+1)^\mu}\,\,\e^{\mu\log2\cdot(k-j)}\,.
$$
We conclude that if $\Spec(H)$ satisfies (B.1) then $\Spec(H_a)$
satisfies (B.3). \qed
\enddemo

\proclaim{Corollary B.3}
Proposition B.1 (modified) implies Proposition B.1.
\endproclaim
\demo{Proof}
Suppose that $\Spec(H)$ satisfies (B.1). Decompose, in accordance with
Lemma B.2, $H=\sum^\oplus_{a\in\BZ_+} H_a$, and apply to each summand
Proposition B.1 (modified) getting this way a family of perturbations
$V_a(t)$, $a\in\BZ_+$ (acting in mutually orthogonal subspaces). Then
the perturbation $\Vt:=\sum^\oplus_{a\in\BZ_+} V_a(t)$ obeys the
conclusions of Proposition B.1. \qed
\enddemo

\demo{Construction of the perturbation}
Set
$$
\Vt:=\sum_{j\in\BN}\sum_{k\in\BN}2\cos([\,\w^{-1}|E_j-E_k|]t)\,
\sqrt{\xi_j\xi_k}\,\,\langle e_k,\cdot\rangle_\HH\,e_j
$$
where
$$
\xi_k:={1\over k\,\log^2(k+1)},\quad k\in\BN.
\tag B.4
$$
Here $[x]$ denotes the integer part of $x\in\BR$. In other words, the
definition of $\Vt$ means that
$$
\langle e_j,\Vt e_k\rangle_\HH=
2\cos([\,\w^{-1}|E_j-E_k|]t)\,\sqrt{\xi_j\xi_k},\quad j,k\in\BN.
$$
Furthermore, from the prescription (\tgV) one finds that $V_{mn}$, with
$m\neq n$, equals either $\sqrt{\xi_{m_2}\xi_{n_2}}$ or 0, and the
former case occurs if and only if
$|m_1-n_1|=[\,\w^{-1}|E_{m_2}-E_{n_2}|]$. For the diagonal entries
we have $V_{mm}=2\xi_{m_2}$.
\enddemo

Before proving that $\Vt$ actually fulfills the conclusions of
Proposition B.1 (modified) we shall derive some auxiliary results.
Nevertheless one can make already now some straightforward observations.
First, $\Vt$ is $2\pi$-periodic and the matrix
$(\langle e_j,\Vt e_k\rangle_\HH)$ is real and symmetric. Second,
$$
\sum_{j\in\BN}\sum_{k\in\BN}|\langle e_j,\Vt e_k\rangle_\HH|^2
\le 4\left(\sum_{k\in\BN}\xi_k\right)^2 <\infty
$$
and so $\Vt$ is Hilbert-Schmidt, for each $t$, and strongly continuous
in $t$. Third, assuming that an index $\eta\in\BZN$ has been chosen,
we find that if $V_{n\eta}\neq0$ and $n_2>\eta_2$
then (again $F\equiv F_\eta$)
$$
F_n-F=\w\{\w^{-1}(E_{n_2}-E_{\eta_2})\}\quad\for\
n_1=\eta_1-[\,\w^{-1}(E_{n_2}-E_{\eta_2})]\,,
$$
and
$$
F_n-F>E_{n_2}-E_{\eta_2}>0\quad\for\
n_1=\eta_1+[\,\w^{-1}(E_{n_2}-E_{\eta_2})]\,.
$$
Here
$$
\{x\}:=x-[x]\in[\,0,1[
$$
is the fractional part of $x\in\BR$ (in the text one has to distinguish
between the fractional part $\{x\}$ and the sequence $\{x_k\}_k$). Hence
$$
\sum_{n\in\BZN,\ n_2>\eta_2}{1\over F_n-F}\,|V_{n\eta}|^2 >
{\xi_{\eta_2}\over\w}\sum_{k\in\BN,\ k>\eta_2}
{\xi_k\over\left\{\w^{-1}\bigl(E_k-E_{\eta_2}\bigr)\right\} } \,.
\tag B.5
$$
Let us add an obvious remark that the sub-sum of (B.2), with the
summation index being restricted by $n_2\le\eta_2$, has only finitely
many nonzero summands.

In the remainder of this appendix we adopt the point of
view of the theory of probability. 
More precisely, the Lebesgue measure on $[\,0,1\,]$ will be
interpreted as a probability measure. This is reflected in the notation,
too. We write, for a measurable set $A\subset[\,0,1\,]$, $\BP(A)$
instead of $|A|$, and consider the measurable functions on the interval
$[\,0,1\,]$ as random variables; here we denote them by the capital
letters $X,Y,Z,\dots$. As usual, $\BE(X)$ means the mathematical
expectation (mean value).

Denote by $\chi_N$, with $N\in[\,1,+\infty[$, the characteristic
function of the interval $]N^{-1},1[$.

\proclaim{Lemma B.4}
Suppose that $M,N\in[\,1,+\infty[$ and $p,q\in\BR$, $1\le p\le q$. Set,
for $\zeta\in\BR$,
$$
\align
Y(\zeta) &:= \sum_{k\in\BZ}\chi_M(p\zeta-k)\,{1\over p\zeta-k}\,, \\
Z(\zeta) &:= \sum_{k\in\BZ}\chi_N(q\zeta-k)\,{1\over q\zeta-k}\,.
\endalign
$$
Then it holds, for the restrictions of $Y$ and $Z$ to the interval
$[\,0,1\,]$, that
$$
|\BE(YZ)-\BE(Y)\BE(Z)|\le 9M\,{p\over q}\,\log N\,.
\tag B.6
$$
\endproclaim
\demo{Proof}
The verification of (B.6) is based on explicit calculations and rather
lengthy but elementary estimates. We only sketch the proof while
indicating some intermediate steps.

Observe that the function $Y$ is $p^{-1}$-periodic and, for each
$k\in\BZ$, it vanishes on the interval $[\,p^{-1}k,p^{-1}(k+M^{-1})\,]$
and is decreasing on $]p^{-1}(k+M^{-1}),p^{-1}(k+1)[$ from the limit
value $M$ to the limit value 1. The integral of $Y$ over the period is
$p^{-1}\log M$ and so it is clear that the mathematical expectation
$\BE(Y)$ is close to $\log M$ provided $p$ is large. More precisely,
$$
|\BE(Y)-\log M|\le{1\over p}\,\log M,\quad
|\BE(Z)-\log N|\le{1\over q}\,\log N\,.
\tag B.7
$$
Let us consider a bit more general situation and compose $Z$ with a
translation,
$$
Z_a(\zeta):=Z(\zeta+q^{-1}a)\quad\text{where}\ a\in\BR.
$$
This time it holds
$$
|\BE(Z_a)-\log N|\le{2\over q}\,\log N\,.
\tag B.8
$$

As a next step we treat the particular case with $p=1$. We claim that
$$
p=1\Longrightarrow
|\BE(YZ_a)-\BE(Y)\BE(Z_a)|\le 3\,{M\over q}\,\log N\,.
\tag B.9
$$
Indeed, now we have the precise equality $\BE(Y)=\log M$ and so
(note that $\log x\le x/\e$ for $x\ge1$)
$$
|\BE(Y)\BE(Z_a)-\log(M)\log(N)|\le{2\over q}\,\log(M)\log(N) <
{M\over q}\,\log N\,.
\tag B.10
$$
Furthermore, it holds
$$
\BE(YZ_a)=\sum_{k(a)\le k<q-N^{-1}+a}\,
\int_{\max\{M^{-1},q^{-1}(k+N^{-1}-a)\}}^{\min\{1,q^{-1}(k+1-a)\}}
\,{d\zeta\over \zeta(q\zeta+a-k)}
$$
where $k(a):=[M^{-1}q+a]$. It follows that
$$
\aligned
\BE(YZ_a)\le &{M\over q}\,\int_{k(a)+N^{-1}-a}^{k(a)+1-a} \,
{d\xi\over \xi+a-k(a)}\qquad + \\
& \sum_{k(a)<k<q-N^{-1}+a}\, {1\over k-a}
\,\int_{k+N^{-1}-a}^{k+1-a}\, {d\xi\over \xi+a-k}
\endaligned
$$
and
$$
\BE(YZ_a)\ge \sum_{k(a)\le k\le q-1+a}\, {1\over k+1-a}\,
\int_{k+N^{-1}-a}^{k+1-a}\, {d\xi\over \xi+a-k} \,.
$$
Proceeding this way one derives rather straightforwardly that
$$
|\BE(YZ_a)-\log(M)\log(N)|\le 2\,{M\over q}\,\log N\,.
\tag B.11
$$
The inequalities (B.10) and (B.11) imply (B.9).

The just discussed particular case (B.9) will be useful when verifying
the general case. Set
$$
J(k):=[\,p^{-1}(k+M^{-1}),p^{-1}(k+1)]\quad\for\
0\le k\le [p]-1\,.
$$
We put also $J([p])$ equal to $[\,p^{-1}([p]+M^{-1}),1\,]$ if
$M^{-1}<\{p\}$ and to $\emptyset$ otherwise. Thus we get
$$
\aligned
|\BE(YZ)-\BE(Y)\BE(Z)| &= |\BE\bigl(Y(Z-\BE(Z))\bigr)| \\
&\le \sum_{0\le k\le[p]}\,
\biggl| \int_{J(k)}{1\over p\zeta-k}\,Z(\zeta)\,d\zeta -
\log N\int_{J(k)} {d\zeta\over p\zeta-k}\,\biggr| \\
&\qquad +\biggl( \,
\sum_{0\le k\le[p]}\int_{J(k)}{d\zeta\over p\zeta-k} \biggr)\,
|\log N-\BE(Z)| \,.
\endaligned
\tag B.12
$$
According to (B.7), the last term in (B.12) can be estimated from
above by
$$
([p]+1)\,{1\over p}\,\log M\cdot {1\over q}\,\log N <
{M\over q}\,\log N\,.
\tag B.13
$$
On the other hand, for $0\le k<[p]$,
$$
\align
\biggl| \int_{J(k)}{1\over p\zeta-k}\,Z(\zeta)\,d\zeta -
\log N\int_{J(k)} {d\zeta\over p\zeta-k}\biggr|
&\le {1\over p}\,
\biggl| \int_{M^{-1}}^1{1\over\xi}\,\tilde Z(\xi)\,d\xi
-\log(M)\BE(\tilde Z) \biggr| \\
&\qquad +{1\over p}\,\log(M)\,|\BE(\tilde Z)-\log N| \\
&\le 4\,{M\over q}\,\log N
\tag B.14
\endalign
$$
where we have set $\tilde Z(\xi):=Z(p^{-1}\xi+p^{-1}k)$ and then we
have applied (B.9) and (B.8) with $\tilde q=q/p$ and $\tilde a=kq/p$.
Quite similarly one estimates
$$
\biggl| \int_{J([p])}{1\over p\zeta-[p]}\,Z(\zeta)\,d\zeta -
\log N\int_{J([p])} {d\zeta\over p\zeta-[p]}\,\biggr|
\le 4\,{M\over q}\,\log N \,.
\tag B.15
$$
Combining (B.12), (B.13), (B.14) and (B.15) we get
$$
|\BE(YZ)-\BE(Y)\BE(Z)|\le ([p]+1)\,4\,{M\over q}\,\log N+
{M\over q}\,\log N\le 9M\,{p\over q}\,\log N\,,
$$
as required. \qed
\enddemo

Next we will treat a sequence of random variables of the same type as
$Y$ but with a particular choice of the parameters $M$ and $p$. Fix
$\theta$, $0<\theta<1/2$, and set
$$
Y_k(\zeta):=\sum_{j\in\BZ_+} \chi_{k^\theta}(h_k\zeta-j)\,
{1\over h_k\zeta-j}\,,\quad k\in\BN,
\tag B.16
$$
where $\{h_k\}_k$ is a sequence of positive numbers. We assume that
$h_k\ge 1$, $\forall k$, and that the sequence obeys the same type of
condition as given in (B.3); this is to say that there exist constants
$C_M>0$ and $\mu>0$ such that
$$
j<k \Longrightarrow {h_k\over h_j}\ge C_M\, \e^{\mu(k-j)}\,.
\tag B.17
$$

Let us specialize some estimates to the random variable $Y_k$. According
to (B.7) we have
$$
|\BE(Y_k)-\theta\log k|\le {\theta\log k\over h_k}\,.
\tag B.18
$$
Quite similarly, it holds true that
$$
|\BE(Y_k^2)-k^\theta+1|\le {k^\theta-1\over h_k} \,.
\tag B.19
$$
As a consequence of (B.19) we get
$$
\BE(Y_k^2)\le 2\,k^\theta\,.
\tag B.20
$$
Finally, Lemma B.4 jointly with (B.17) tells us that
$$
\align
j<k \Longrightarrow |\BE(Y_j Y_k)-\BE(Y_j)\BE(Y_k)| &\le
9j^\theta\,\log(k^\theta)\,{h_j\over h_k} \\
&\le {9\theta\over C_M}\,j^\theta\,\log(k)\,\e^{-\mu(k-j)}\,.
\tag B.21
\endalign
$$

\proclaim{Lemma B.5}
Suppose that a sequence $\{h_k\}_{k\in\BN}$ satisfies $h_k\ge 1$,
$\forall k$, and the condition (B.17). Set
$$
S_N:=Y_1+\dots+Y_N,\quad N\in\BN,
\tag B.22
$$
where $Y_k$ has been defined in (B.16). Then
$$
\lim_{N\to\infty}{1\over N}\,\bigl(S_N-\BE(S_N)\bigr)=0
\tag B.23
$$
almost everywhere on $[\,0,1\,]$.
\endproclaim
\demo{Proof}
This would be a classical text-book result if the random variables $Y_k$
were independent (see \S5.1 in \cite{\Chung}). The estimate (B.21)
guarantees that, in our case, the random variables are correlated
sufficiently weakly. We only sketch the proof. Set
$$
Y'_k:=Y_k-\BE(Y_k),\quad S'_N:=S_N-\BE(S_N)=Y'_1+\dots+Y'_N\,.
$$
Fix $\theta'$ such that $\theta<\theta'<1/2$. Using (B.20) and (B.21) we
estimate
$$
\align
\BE\bigl((S'_N)^2\bigr) &= \sum_{1\le k\le N}\BE\bigl((Y'_k)^2\bigr)
+2\sum_{1\le j< k\le N}  \BE(Y'_jY'_k) \\
&\le \sum_{1\le k\le N} \BE(Y_k^2)+
2\sum_{1\le j<k\le N} |\BE(Y_jY_k)-\BE(Y_j)\BE(Y_k)| \\
&\le 2 \sum_{1\le k\le N} k^\theta +{18\theta\over C_M}
\sum_{1\le j\le N} j^\theta\log N\sum_{s\in\BN} e^{-\mu s} \\
&\le C_I\,N^{1+\theta'}
\endalign
$$
where $C_I>0$ is a constant. According to the Chebyshev's inequality we
have, for $\veps>0$,
$$
\BP(|S'_N|>N\veps) \le {C_I\,N^{1+\theta'}\over N^2\veps^2}
={C_I\over \veps^2N^{1-\theta'}}\,,
$$
and so
$$
\sum_{N\in\BN} \BP(|S'_N|>N^2\veps)<\infty\,.
$$
By the Borel-Cantelli lemma, $\limsup N^{-2}|S'_{N^2}|\le\veps$ holds
true for all $\veps>0$ and for almost all $\zeta\in[\,0,1\,]$.
One deduces from this, by a standard argument, that
$$
\lim_{N\to\infty}{1\over N^2} S'_{N^2}\equiv
\lim_{N\to\infty}{1\over N^2}\bigl(S_{N^2}-\BE(S_{N^2})\bigr)=0
\tag B.24
$$
almost everywhere on $[\,0,1\,]$. To pass from (B.24) to (B.23) we
introduce the random variables
$$
D_N:=\max_{N^2<M<(N+1)^2}|S'_{N^2}-S'_M|\,.
$$
Using basically the same estimates as before one shows that
$$
\sum_{N\in\BN}\BP(D_N>N^2\veps)\le
\sum_{N\in\BN} {C_{II}\over \veps^2 N^{2(1-\theta')}}<\infty
$$
and so $\lim N^{-2} D_N=0$ almost everywhere on $[\,0,1\,]$. To complete
the proof it suffices to observe that, for $N^2<M<(N+1)^2$,
$$
{|S'_M|\over M}\le {|S'_{N^2}|+D_N\over N^2}\,.\quad \qed
$$
\enddemo

\proclaim{Lemma B.6}
Suppose that $\{h_k\}_{k\in\BN}$, a sequence of positive numbers,
satisfies the condition (B.17),
and $\xi_k$ is the same as in (B.4). Then
$$
\sum_{k\in\BN}{\xi_k\over \{\zeta h_k\} }=+\infty
\tag B.25
$$
for almost all $\zeta>0$.
\endproclaim
\demo{Proof}
Clearly it suffices to show that (B.25) is valid for almost all $\zeta$
from an arbitrary bounded interval $[\,0,z\,]$, $z>0$. Having observed
that the condition (B.17) is invariant with respect to the scaling of
$h_k$ we can restrict ourselves to $\zeta\in[\,0,1\,]$. Furthermore, the
conclusion of the lemma is not influenced by omitting several first
numbers of the sequence $\{h_k\}_k$. This is why we can assume that the
assumptions of Lemma B.5 are satisfied.

Set
$$
X_k(\zeta):=1/\{\zeta h_k\},\quad k\in\BN .
$$
Note that $Y_k$ (cf. (B.16)) is nothing but the cutoff of the function
$X_k$ obtained by annulling the values which exceed the level
$k^\theta$; hence $X_k\ge Y_k$. The symbol $S_N$ retains its meaning
from (B.22). We have
$$
\align
\sum_{k\in\BN} \xi_k X_k &\ge \sum_{k\in\BN} \xi_k Y_k \\
&= \sum_{k\in\BN} (\xi_k-\xi_{k+1})(S_k-\BE(S_k))+
\sum_{k\in\BN} (\xi_k-\xi_{k+1})\BE(S_k) \,.
\endalign
$$
It is elementary to derive the estimate
$$
{1\over (1+k)^2\log^2(2+k)}\le \xi_k-\xi_{k+1} \le
{3\over k^2\log^2(1+k)} \,.
$$
Hence $\sum_k k(\xi_k-\xi_{k+1})<\infty$. Since, by Lemma B.5, the
sequence $\{k^{-1}(S_k-\BE(S_k))\}_{k\in\BN}$ is bounded almost
everywhere we find that the sum
$$
\sum_{k\in\BN} (\xi_k-\xi_{k+1})(S_k-\BE(S_k))
$$
converges for almost all $\zeta$. To finish the proof we need to
estimate $\BE(S_k)$. The inequality (B.18) and the fact that
$\lim h_k=+\infty$ imply that there exists $k_\star\in\BN$ such that
$$
\BE(Y_k)\ge\half\,\theta\log k,\quad\for\ k\ge k_\star\,.
$$
Consequently, if $k\ge k_\star$, then
$$
\BE(S_k)\ge \half\,\theta\sum_{k_\star\le j\le k}\log j \ge
C_{III}\,(1+k)\log(2+k)
$$
where $C_{III}>0$ is a constant. Hence
$$
\sum_{k\in\BN}(\xi_k-\xi_{k+1})\BE(S_k)\ge
C_{III}\sum_{k\ge k_\star}{1\over (1+k)\log(2+k)}=+\infty \,.
$$
This completes the proof. \qed
\enddemo

\demo{Proof of Proposition B.1 (modified)}
In virtue of the inequality (B.5) and the remark following it, it is
sufficient to apply Lemma B.6 to the sequence $\{h_k\}_{k\in\BN}$
defined by
$$
h_k:=E_k-E_{\eta_2},\quad k\in\BN,\ k>\eta_2 \,,
$$
(in fact, we treat a countable family of such sequences labeled by
$\eta_2\in\BN$). Observe that if
$\eta_2<j<k$ then $h_k/h_j\ge E_k/E_j$ and so (B.3) implies (B.17).
Hence the assumption of Lemma B.6 is indeed satisfied. \qed
\enddemo

\bigpagebreak
\flushpar {\bf Appendix C. Lipschitz functions}
\medpagebreak

Here we present some auxiliary results concerning Lipschitz functions
which are quite straightforward to verify but are not mentioned in
\cite{\Stein}, our main source on this subject. Moreover, in
view of applications we are interested in, we allow the target space
to be generally a Banach algebra (sometimes only a Banach space) rather
than $\BC$. In fact, this doesn't cause any essential complication --
one has just to be careful about the order of multipliers in all
expressions. The notation in this appendix is autonomous, particularly
the symbols $f,g,P$ etc. have different meaning in the main text of the
paper.

\proclaim{Definition C.1}
Suppose that $\Pi\subset\BR^n$ is a closed set and $\AA$ is a Banach
algebra (or just a Banach space). A function $f$ defined on $\Pi$ and
with values in $\AA$ belongs to the Lipschitz class
$\Lip(\ell+\veps,\Pi)$, with $\ell\in\BZ_+$ and $0<\veps\le1$, if and
only if there exists a family of functions
$\{f\en ;\ \nu\in\BZ_+^n,\ |\nu|\le\ell\}$, with $f^{(0)}\equiv f$,
and a constant $M>0$ such that, for all $\nu\in\BZ_+^n$,
$|\nu|\le\ell$, it holds true that
$$
\align
& |f\en(x)|\le M,\quad\fora\ x\in\Pi, \\
& |f\en(x)-\pr_x^\nu\,P(x,y)|\le M\,|x-y|^{\ell+\veps-|\nu|},\quad
\fora\ x,y\in\Pi,
\endalign
$$
where
$$
P(x,y):=\sum_{\mu,\ |\mu|\le\ell} f\em(y)\,{(x-y)^\mu\over\mu!}\,.
$$
The smallest constant $M$ with this property is called the \Lp norm
$M(f)$.
\endproclaim

As one can guess, we have denoted the norm in $\AA$ by $|\cdot|$. If not
specified otherwise, the multiindices $\mu,\nu,\dots$ are assumed to
belong to $\BZ_+^n$. We use the partial ordering on $\BZ_+^n$:
$\mu\le\nu$ means that $\mu_j\le\nu_j$ for all $j$, $1\le j\le n$. Set
$$
R_\nu(x,y):=f\en(x)-\pr_x^\nu\,P(x,y),\quad\nu\in\BZ_+^n,\ |\nu|\le\ell.
$$
If necessary, the dependence of $P(x,y)$ or $R(x,y)$ on $f$ will be
distinguished by a superscript. A detailed proof of the following basic
theorem is given in \cite{\Stein}.

\proclaim{Theorem C.2 (Whitney Extension Theorem)}
There exists a continuous mapping
$\EE:\Lip(\ell+\veps,\Pi)\to\Lip(\ell+\veps,\BR^n)$ such that
$\EE(f)$ is an extension of $f$ for all $f\in\Lip(\ell+\veps,\Pi)$.
The norm of $\EE$ has a bound independent of $\Pi$.
\endproclaim

We shall frequently use the observation that
$f\in\Lip(\ell+\veps,\BR^n)$ if and only if $f\in C^\ell(\BR^n)$,
all derivatives of $f$ up to the order $\ell$ are uniformly bounded on
$\BR^n$, and
$\pr_x^\nu\,f\in\Lip(\veps,\BR^n)$ for all $\nu$, $|\nu|=\ell$. This
claim still holds true when replacing $\BR^n$ by a closed convex
subset $\Pi$ of dimension $n$. Clearly,
$$
f\in\Lip(\ell+\veps,\Pi)\Longrightarrow
f\en\in\Lip(\ell-|\nu|+\veps,\Pi),\quad\fora\ \nu,\ |\nu|\le\ell.
\tag C.1
$$
The family of functions corresponding to $f\en$ is
$\{f^{(\nu+\mu)}\}_{0\le|\mu|\le\ell-|\nu|}$. The extension operator
has the property that $f\en(x)=\pr_x^\nu\,\EE(f)(x)$ holds true for
all $x\in\Pi$ and all $\nu\in\BZ_+^n$, $|\nu|\le\ell$. The
following proposition is quite easy to verify.

\proclaim{Proposition C.3}
Suppose that $\ell\ge1$ and $\Pi$ is bounded. If
$f\in\Lip(\ell+\veps,\Pi)$ then
$f\in\Lip(\ell'+\veps',\Pi)$ for all $\ell'$, $0\le\ell'<\ell$,
and any $\veps'$, $0<\veps'\le 1$. The embedding mapping
$\II_{\ell,\ell'}:\Lip(\ell+\veps,\Pi)\to\Lip(\ell'+\veps',\Pi)$,
sending the family $\{f\en\}_{|\nu|\le\ell}$ to
$\{f\en\}_{|\nu|\le\ell'}$, is bounded.
\endproclaim

In the following two propositions we shall need the
structure of algebra on $\AA$.

\proclaim{Proposition C.4}
Suppose that $\Pi$ is bounded and both $f$ and $g$ belong to
$\Lip(\ell+\veps,\Pi)$. Then $fg\in\Lip(\ell+\veps,\Pi)$.
\endproclaim
\demo{Proof}
Set $h=fg$ and, more generally,
$$
h\en=\sum_{\mu,\ \mu\le\nu}{\nu!\over\mu!(\nu-\mu)!}\,f\em
g^{(\nu-\mu)} \,.
$$
Then
$$
P^h(x,y)=P^f(x,y)\,P^g(x,y)-\Psi(x,y)
$$
where
$$
\Psi(x,y):=\sum\Sb \mu,\nu \\
|\mu|\le\ell,\ |\nu|\le\ell,\ |\mu+\nu|\ge\ell+1 \endSb \,\,
f\em(y)\,g\en(y)\,{(x-y)^{\mu+\nu}\over \mu!\,\nu!} \,.
\tag C.2
$$
It follows that
$$
\align
h\en(x) & -\pr_x^\nu\,P^h(x,y) =\\
\sum_{\mu,\ \mu\le\nu}
{\nu!\over\mu!(\nu-\mu)!}\, \bigl(R^f_\mu(x,y)\,
\pr_x^{\nu-\mu}\,P^g(x,y)+f\em(x)\,R^g_{\nu-\mu}(x,y)\bigr) \\
&\qquad +\pr_x^\nu\,\Psi(x,y)\,.
\endalign
$$
We conclude that $|h\en(x)|=O(1)$ and
$|R^h_\nu(x,y)|=O(|x-y|^{\ell+\veps-|\nu|})$. \qed
\enddemo

\proclaim{Proposition C.5}
Suppose that $\Pi$ is  bounded, $f\in\Lip(\ell+\veps,\Pi)$, $f(x)^{-1}$
exists in $\AA$ for all $x\in\Pi$, and $|f(x)^{-1}|$ is uniformly
bounded on $\Pi$. Then $g\in\Lip(\ell+\veps,\Pi)$ where
$g(x)=f(x)^{-1}$.

If, in addition, the diameter $\diam\Pi\le1$ and $|f(x)^{-1}|\le\kappa$
for all $x\in\Pi$ then
$$
M(g)\le C_L\,M(f)^{\ell+1}\kappa^{\ell+2}
$$
where $C_L\equiv C_L(n,\ell)$ is a constant.
\endproclaim
\demo{Proof}
We can assume from the beginning that $\diam\Pi\le1$ and, by rescaling
$f$, that $|f(x)^{-1}|\le1$ on $\Pi$, i.e., $\kappa=1$ (the norm
$M(\cdot)$ is homogeneous). Then we have
$$
|f(x)|\ge |f(x)^{-1}| |f(x)| \ge 1
$$
and so $M(f)\ge 1$.

We define successively, for $1\le|\nu|\le\ell$,
$$
g\en=-g\biggl(\,\sum_{\mu,\ \mu<\nu}{\nu!\over\mu!(\nu-\mu)!}\,
f^{(\nu-\mu)}g\em\biggr) \,.
\tag C.3
$$
This means that the identity
$$
\sum_{\mu,\ \mu\le\nu}{\nu!\over\mu!(\nu-\mu)!}\,
f^{(\nu-\mu)}(x)\,g\em(x)=\delta_{\nu 0}
\tag C.4
$$
is valid for all $\nu\in\BZ_+^n$, $|\nu|\le\ell$, and all $x\in\Pi$.

Clearly, all $g\en$ are bounded. We know, by the assumption, that
$|g(x)|\le1$, and we claim that
$$
|g\en(x)|\le\bigl(|\nu|\,M(f)\bigr)^{|\nu|}\quad\for\
1\le|\nu|\le\ell,\ \forall x\in\Pi.
\tag C.5
$$
To see (C.5) we proceed by induction $|\nu|$ using the formula (C.3),
$$
\align
|g\en(x)| &\le \sum_{\mu,\ \mu<\nu}{\nu!\over\mu!(\nu-\mu)!}\,
M(f)\,\bigl(|\mu|\,M(f)\bigr)^{|\mu|} \\
&\le M(f)^{|\nu|}\sum_{\mu,\ \mu\le\nu}{\nu!\over\mu!(\nu-\mu)!}\,
(|\nu|-1)^{|\mu|} \\
&= \bigl(|\nu|\,M(f)\bigr)^{|\nu|} \,.
\endalign
$$

With the aid of (C.4) one finds readily that
$$
P^f(x,y)\,P^g(x,y)=1+\Psi(x,y)
\tag C.6
$$
where $\Psi$ is the same as in (C.2). Differentiating (C.6) and
subtracting (C.4) from the result one arrives at ($0\le|\nu|\le\ell$)
$$
\sum_{\mu,\ \mu\le\nu}{\nu!\over\mu!(\nu-\mu)!}\,
\bigl(\pr_x^{\nu-\mu}\,P^f(x,y)\cdot\pr_x^\mu\,P^g(x,y)-
f^{(\nu-\mu)}(x)\,g\em(x)\bigl)=\pr_x^\nu\,\Psi(x,y)\,.
\tag C.7
$$
More conveniently, let us rewrite (C.7) as a recurrence formula,
$$
\aligned
-f(x)\,R^g_\nu(x,y) &=
\sum_{\mu,\ \mu<\nu}{\nu!\over\mu!(\nu-\mu)!}\,
f^{(\nu-\mu)}(x)\,R^g_\mu(x,y) \\
&\qquad +\sum_{\mu,\ \mu\le\nu}{\nu!\over\mu!(\nu-\mu)!}\,
R^f_{\nu-\mu}(x,y)\,\pr_x^\mu\,P^g(x,y) \\
&\qquad +\pr_x^\nu\,\Psi(x,y)\,.
\endaligned
\tag C.8
$$

Using (C.5) one finds easily the required bounds for a part of the RHS
of (C.8). Namely,
$$
\align
|\pr_x^\nu\,P^g(x,y)| &\le \sum_{\mu,\ |\mu|\le\ell-|\nu|}
\bigl(|\nu+\mu|\,M(f)\bigr)^{|\nu+\mu|}\,{|x-y|^{|\mu|}\over\mu!} \\
&\le \sum_{\mu\in\BZ_+^n}{1\over\mu!}\,\bigl(\ell M(f)\bigr)^\ell
\endalign
$$
and so ($|x-y|\le1$)
$$
|R^f_{\nu-\mu}(x,y)\,\pr_x^\mu\,P^g(x,y)|\le \e^n\ell^\ell
M(f)^{\ell+1}|x-y|^{\ell+\veps-|\nu|}\,.
\tag C.9
$$
Furthermore,
$$
\align
|\pr_x^\nu\Psi(x,y)| &\le \sum\Sb |\sg|\le\ell,\ |\mu|\le\ell \\
|\sg+\mu|\ge\ell+1,\ \sg+\mu\ge\nu \endSb \,
|f^{(\sg)}(y)|\,|g\em(y)|\,
\frac{(\sg+\mu)!}{\sg!\,\mu!\,(\sg+\mu-\nu)!}\,
|x-y|^{|\sg+\mu|-|\nu|} \\
&\le \const\ M(f)^{\ell+1}\,|x-y|^{\ell+1-|\nu|} \,.
\tag C.10
\endalign
$$
So it remains to estimate, in the same manner, the first sum on the RHS
of (C.8). The three estimates ((C.9), (C.10) and the one still lacking)
should amount in the existence of constants
$c_\nu>0$ (depending also on $n$ and $\ell$, $0\le|\nu|\le\ell$) such
that
$$
|R^g_\nu(x,y)|\le c_\nu\,M(f)^{\ell+1}|x-y|^{\ell+\veps-|\nu|}\,.
\tag C.11
$$

To prove the proposition we shall proceed by induction
in $\ell$. The case $\ell=0$ is obvious for $|g(x)|\le 1\le M(f)$ and
$$
|R^g(x,y)|=|-f(x)^{-1}R^f(x,y)\,f(y)^{-1}|\le M(f)\,|x-y|^\veps\,;
$$
hence $M(g)\le M(f)$.
Suppose now that $\ell\ge1$ and the proposition is valid for all
$\ell'$, $0\le\ell'<\ell$. Write, for $\mu<\nu$,
$$
R^g_\mu(x,y)=R'_\mu(x,y)+
\sum\Sb \sg \\ \ell'-|\mu|<\sg\le\ell-|\mu| \endSb
g^{(\mu+\sg)}(y)\,{(x-y)^\sg\over \sg!}
$$
where $\ell'=\ell-|\nu-\mu|$ and $R'_\mu$ is the rest function related
to $\II_{\ell,\ell'}(g)\in\Lip(\ell'+1,\Pi)$. Note that
$|\mu|\le\ell'<\ell$. By the induction hypothesis and by
Proposition C.3, we have ($\ell'+1-|\mu|\ge\ell+\veps-|\nu|$)
$$
\align
|R'_\mu(x,y)| &\le C_L(n,\ell')\,M(\II_{\ell,\ell'}(f))^{\ell'+1}
|x-y|^{\ell'+1-|\mu|} \\
&\le C_L(n,\ell')\,\|\II_{\ell,\ell'}\|^\ell\,M(f)^\ell
|x-y|^{\ell+\veps-|\nu|} \,.
\endalign
$$
In addition, for $|\sg|>\ell'-|\mu|$ and $|\mu+\sg|\le\ell$,
$$
|g^{(\mu+\sg)}(y)\,(x-y)^\sg|\le \bigl(\ell\,M(f)\bigr)^\ell
|x-y|^{\ell'+1-|\mu|}\le
\bigl(\ell\,M(f)\bigr)^\ell |x-y|^{\ell+\veps-|\nu|} \,.
$$
We conclude that
$$
|f^{(\nu-\mu)}(x)\,R^g_\mu(x,y)|\le \const\, M(f)^{\ell+1}
|x-y|^{\ell+\veps-|\nu|} \,.
\tag C.12
$$
The formula (C.8) and the bounds (C.9), (C.10) and (C.12) prove the
validity of (C.11). \qed
\enddemo

The last auxiliary result concerns the composition of functions. This
time $\AA$ is a Banach space.

\proclaim{Proposition C.6}
Suppose that $g:\BR^n\to\AA$ belongs to $\Lip(\ell+\veps,\BR^n)$.
\roster 
\item"(i)" 
If $\ell=0$ and $f:\Pi\to\BR^n$ belongs to $\Lip(1,\Pi)$ then
$g\circ f\in\Lip(\veps,\Pi)$.
\item"(ii)"
If $\ell\ge1$ and $f:\BR^m\to\BR^n$ belongs to $\Lip(\ell+\veps,\BR^m)$
then $g\circ f\in\Lip(\ell+\veps,\Pi)$ for any compact set
$\Pi\subset\BR^m$.
\endroster
\endproclaim  
\demo{Proof}      
(i) This is obvious from the estimates $|g\circ f(x)|\le M(g)$ and
$$
|g\circ f(x)-g\circ f(y)|\le M(g)\,|f(x)-f(y)|^\veps\le
M(g)\,M(f)^\veps |x-y|^\veps \,.
$$

(ii) We can restrict ourselves to the case when $\Pi=\bar\UU$ where
$\UU\subset\BR^m$ is an open, convex and bounded set. Write $f$ as an
$n$-tuple of functions: $f=(f_1,\dots,f_n)$, $f_j:\BR^m\to\BR$. Clearly
$g\circ f\in C^\ell(\Pi)$, $\Pi$ is compact, and thus we have to show
only that
$\pr_x^\nu\,g\circ f\in\Lip(\veps,\Pi)$ for all $\nu$, $|\nu|=\ell$.
However, $\pr_x^\nu\,g\circ f$ is a polynomial in $\pr_x^\mu\,f_j$,
$1\le j\le n$ and $|\mu|\le\ell$, and in $(\pr_y^\mu\,g)\circ f$,
$|\mu|\le\ell$. This means that, when applying an obvious modification
of Proposition C.4 (here we multiply scalar functions by
vector-valued functions), it suffices to verify that all the
multipliers belong to $\Lip(\veps,\Pi)$. By Proposition C.3,
$\pr_x^\mu\,f_j\in\Lip(\veps,\Pi)$ and $f\in\Lip(1,\Pi)$. Furthermore,
from the already proven part (i) we conclude that
$(\pr_y^\mu\,g)\circ f\in\Lip(\veps,\Pi)$. This completes the proof.
\qed
\enddemo

\vskip 12pt
\noindent{\bf Acknowledgements.} P. S. wishes to thank his hosts at
Centre de Physique Th\'eorique in Marseille and at Universit\'e de
Toulon et du Var for their hospitality. The support from the grant
202/96/0218 of Czech Grant Agency is also gratefully acknowledged.


\vskip 0.3in

\Refs
\widestnumber\key{13}

\vskip 0.1in

\ref\key 1\by Arnold, V.I.
\paper Small Divisors II. Proof of the A.N.~Kolmogorov Theorem on
conservation of conditionally periodic motions under small perturbations of
the Hamiltonian function
\jour Usp. Mat. Nauk
\vol 18
\yr 1963
\pages  13-40
\endref

\ref\key 2\by Bellissard, J.
\paper Stability and instability in quantum mechanics
\inbook Trends and Developments in the Eighties
\ed Albeverio and Blanchard
\publaddr Singapore
\publ Word Scientific
\yr 1985 \page 1-106
\endref

\ref\key 3\by Bleher, P. M., Jauslin, H. R., Lebowitz J. L.
\paper Floquet spectrum for two-level systems in quasi-periodic
time dependent fields
\jour J. Stat. Phys.
\vol 68
\yr 1992
\pages 271
\endref

\ref\key 4\by Chung, K. L.
\book A course in probability theory
\publ Academic Press
\yr 1970
\endref

\ref\key 5\by Combescure, M.
\paper The quantum stability problem for time-periodic
perturbations of the harmonic oscillator
\jour Ann. Inst. Henri Poincar\'e
\vol 47
\yr 1987
\pages 62-82
\moreref
\paper Erratum
\jour Ann. Inst. Henri Poincar\'e
\vol 47
\yr 1987
\pages 451-454
\endref

\ref\key 6\by Duclos, P., \v S\v tov\'\i\v cek, P.
\paper Floquet Hamiltonians with pure point spectrum
\jour Commun. Math. Phys.
\vol 177
\yr 1996
\pages 327-347
\endref

\ref\key 7\by Duclos, P., \v S\v tov\'\i\v cek, P., Vittot, M.
\paper Perturbation of an eigen-value from a dense point spectrum: an
example
\jour J. Phys. A: Math. Gen.
\vol 30
\yr 1997
\pages 7167-7185
\endref

\ref\key 8\by Eliasson, L. H.
\paper Perturbations of stable invariant tori for Hamiltonian systems
\jour Ann. Scuola Norm. Sup. Pisa Cl. Sci.
\vol IV(15)
\yr 1988
\pages 115-147
\endref

\ref\key 9\by Enss, V., Veseli\'c, K.
\paper Bound states and propagating states for time-dependent
Hamiltonians
\jour Ann. Inst. Henri Poincar\'e
\vol 39
\yr 1983
\pages 159-191
\endref

\ref\key 10\by Howland, J. S.
\paper Scattering theory for Hamiltonians periodic in time
\jour Indiana J. Math.
\vol 28
\yr 1979
\pages 471-494
\endref

\ref\key 11\by Howland, J. S.
\paper Floquet operators with singular spectrum I
\jour Ann. Inst. Henri Poincar\'e
\vol 49
\yr 1989
\pages 309-323
\moreref
\paper Floquet operators with singular spectrum II
\jour Ann. Inst. Henri Poincar\'e
\vol 49
\yr 1989
\pages  325-334
\endref

\ref\key 12\by Joye, A.
\paper Absence of absolutely continuous spectrum of Floquet operators
\jour J. Stat. Phys.
\vol 75
\yr  1994
\pages  929-952
\endref

\ref\key 13\by Kato, T.
\book Perturbation Theory of Linear Operators
\publaddr New York
\publ Springer-Verlag
\yr 1966
\endref

\ref\key 14\by Kolmogorov, A. N.
\paper On the conservation of conditionally periodic motions
under small perturbations of the Hamiltonian function
\jour Dokl. Akad. Nauk. SSSR
\vol 98
\yr  1954
\pages  527-530
\endref

\ref\key 15\by Messiah, A.
\book M\'ecanique Quantique II
\publaddr Paris
\publ Dunod
\yr 1964
\endref

\ref\key 16\by Moser, J.
\paper On invariant curves of area preserving mappings of an
annulus
\jour Nachr. Akad. Wiss. G\"ottingen, II. Math. Phys. Kl
\vol  11a
\yr 1962
\pages 1-20
\endref

\ref\key 17\by Nenciu, G.
\paper Floquet operators without absolutely continuous spectrum
\jour Ann. Inst. Henri Poincar\'e
\vol 59
\yr 1993
\pages 91-97
\endref

\ref\key 18\by De Oliviera, C. R., Guarneri, I., Casati, G.
\paper From power-localization to extended quasi-energy eigenstates in a
quantum periodically driven system
\jour Europhys. Lett.
\vol 27
\yr 1994
\pages 187-192
\endref

\ref\key 19\by Reed, M., Simon, B.
\book Methods of Modern Mathematical Physics IV
\publaddr New York
\publ Academic Press
\yr 1978
\endref

\ref\key 20\by Rellich, F.
\paper St\"orungstheorie der Spektralzerlegung, I.
\jour Math. Ann.
\vol 113
\yr 1937
\pages 600-619
\endref

\ref\key 21\by Sambe, H.
\paper
\jour Phys. Rev. A
\vol 7
\yr 1973
\pages 2203
\endref

\ref\key 21\by Siegel, C. L.
\paper Iterations of analytic functions
\jour Ann. Math.
\vol 43
\yr 1942
\pages 607-612
\endref

\ref\key 22\by Stein, E. M.
\book Singular Integrals and Differentiability Properties of
Functions
\publaddr New Jersey
\publ Princeton University Press
\yr 1970
\endref

\ref\key  23\by Yajima, K.
\paper Scattering theory for Schr\"odinger equations with potential
periodic in time
\jour J. Math. Soc. Japan
\vol 29
\yr 1977
\pages 729-743
\endref

\endRefs
\enddocument